\documentclass[a4paper,english]{article}

\usepackage{amsmath}
\usepackage{amssymb}
\usepackage{graphicx}
\usepackage{amssymb}
\usepackage{ae, aecompl}
\usepackage{graphicx}% Include figure files
\usepackage{dcolumn}% Align table columns on decimal point
\usepackage{bm}% bold math
\usepackage{amssymb}
\usepackage{graphics}
\usepackage{epsfig}
\usepackage[T1]{fontenc}
\usepackage[latin1]{inputenc}
\usepackage{babel}
\usepackage{slashed}
\usepackage{color}

\definecolor{blue}{rgb}{0,0,0.5}
\definecolor{lightblue}{rgb}{0,0,1}
\definecolor{red}{rgb}{0.5,0,0}
\definecolor{lightred}{rgb}{1,0.5,0}
\definecolor{green}{rgb}{0,0.5,0}
\definecolor{darkgreen}{rgb}{0.0,0.3,0.0}
\definecolor{orange}{rgb}{1,0.4,0}
\definecolor{grey}{rgb}{0.5,0.5,0.5}

\providecommand{\higgs}{\phi}

\providecommand{\lepton}{{\ell}}

\providecommand{\majneutrino}{N}

\providecommand{\yu}{h}
\providecommand{\pL}{{S_\ell}}
\providecommand{\pH}{{\Delta_\higgs}}

\providecommand{\pN}{{S}}

\providecommand{\pLH}{{S_{\ell\phi}}}
\providecommand{\pLcp}{{\mbox{$\bar{S}$}_\ell}}
\providecommand{\pHcp}{{\mbox{$\bar{\Delta}_\higgs$}}}
\providecommand{\pNcp}{{\mbox{$\bar{S}$}}}
\providecommand{\pLHcp}{{\mbox{$\bar{S}$}_{\ell\phi}}}
\newcommand{\GammaLH}{\Gamma_{\lepton\higgs}}

\providecommand{\seL}{{\Sigma_\ell}}

\providecommand{\seN}{{\Sigma_N}}

\providecommand{\seLcp}{{\mbox{$\bar{\Sigma}$}_\ell}}

\providecommand{\seNcp}{{\mbox{$\bar{\Sigma}$}_N}}

\newcommand{\La}{\alpha}
\newcommand{\Lb}{\beta}
\newcommand{\Lc}{\gamma}

\newcommand{\Da}{\mu}
\newcommand{\Db}{\nu}
\newcommand{\Dc}{\lambda}
\newcommand{\Dd}{\sigma}

\providecommand{\CTP}{{\cal C}}

\providecommand{\feyn}[3]{\raisebox{#1}{\includegraphics[width=#2\textwidth,keepaspectratio]{#3}}}

\providecommand{\sgn}{\mathrm{sgn}}

\providecommand{\tr}{{\text{tr}}}

\renewcommand{\Re}{\mbox{Re}}
\renewcommand{\Im}{\mbox{Im}}

\newcommand{\eq}{{\!\!\!\! eq}}

\providecommand{\ifeqthenelse}[4]{\edef\tempa{#1}\def\tempb{#2}\ifx\tempa\tempb {#3} \else {#4}\fi}
\providecommand{\feyn}[3]{\raisebox{#1}{\includegraphics[width=#2\textwidth,keepaspectratio]{Feyn/#3}}}

\renewcommand{\vec}[1]{{\mathbf #1}}
\newcommand{\kasten}[1]{#1}
\newcommand{\CP}{\textit{CP}}

\begin{document}

%\date{\mbox{ }}
\date{\today}

\title{
%{\normalsize
%DESY 11-xxx\hfill\mbox{}\\}
\vspace{1.5cm}
\bf Leptogenesis from first principles\\ in the resonant regime\\[8mm]}

\author{Mathias Garny$^a$, Alexander Kartavtsev$^b$, Andreas Hohenegger$^c$\\[2mm]
{\normalsize\it a  Deutsches Elektronen-Synchrotron DESY,}\\[-0.05cm]
{\it\normalsize Notkestra\ss{}e 85, 22603 Hamburg, Germany}\\[2mm]
{\normalsize\it b Max-Planck-Institut f\"ur Kernphysik,}\\[-0.05cm]
{\it\normalsize Saupfercheckweg 1, 69117 Heidelberg, Germany}\\[2mm]
{\normalsize\it c \'Ecole Polytechnique F\'ed\'erale de Lausanne}\\[-0.05cm]
{\it\normalsize CH-1015 Lausanne, Switzerland}}

\maketitle

\thispagestyle{empty}

\begin{abstract}
\noindent
The lepton asymmetry generated by the out-of-equilibrium decays
of heavy Majorana neutrinos with a quasi-degenerate mass spectrum is resonantly enhanced.
In this work, we study this scenario within a first-principle
approach. The quantum field theoretical treatment is applicable for
mass splittings of the order of the width of the Majorana
neutrinos, for which the enhancement is maximally large.
The non-equilibrium evolution of the mixing Majorana neutrino fields
is described by a formal analytical solution of the Kadanoff-Baym equations,
that is obtained by neglecting the back-reaction.
Based on this solution, we derive approximate analytical
expressions for the generated asymmetry and compare them to the Boltzmann
result. We find that the resonant enhancement obtained from the Kadanoff-Baym approach
is smaller compared to the Boltzmann approach, due to additional contributions that
describe coherent transitions between the Majorana neutrino species. We also
discuss corrections to the masses and widths of the degenerate pair of Majorana neutrinos
that are relevant for very small mass splitting, and compare the approximate analytical
result for the lepton asymmetry with numerical results.
\end{abstract}

\newpage

\section{Introduction}

The almost complete absence of  antimatter on Earth, in the solar system and 
in hadronic cosmic rays suggests that the universe is baryonically asymmetric. 
This conclusion is confirmed by experimental data on the abundances of the light 
elements \cite{Kolb:1990vq} and precise measurements of the cosmic
microwave background spectrum \cite{Hinshaw:2008kr,Komatsu:2008hk}.
The baryon asymmetry of the universe can be generated dynamically provided
the three Sakharov conditions~\cite{Sakharov:1967dj} are fulfilled in the early 
universe: violation of baryon (or baryon minus lepton) number; violation of 
\textit{C} and \CP; and deviation from thermal equilibrium. 
Although the Sakharov conditions are fulfilled 
in the  Standard Model of particle physics (SM), the smallness of the 
\CP-violation in the quark sector and the fact that, given the current bounds 
on the Higgs mass, 
the electroweak  transition is not of  first order, do not allow for the generation 
of an asymmetry comparable to the observed one. Therefore, the question about the 
origin of this asymmetry represents a major puzzle of modern physics. 

From the theoretical point of view  a very attractive explanation of
the observed baryon asymmetry of the Universe is provided by the 
baryogenesis via leptogenesis scenario \cite{Fukugita:1986hr}. In this 
scenario the Standard Model is supplemented by at least two right-handed 
Majorana neutrinos $N_i$ which couple to leptons and the Higgs: 
\begin{align}
	\label{lagrangian}
	{\cal L}={\cal L}_{SM}+{\textstyle\frac12}\bar N_i
	\bigl(i\slashed{\partial} -  M_i\bigr)N_i 
	- h_{\alpha i}\bar \ell_\alpha {\tilde \phi}  P_R N_i
	-h^\dagger_{i\alpha} \bar N_i {\tilde \phi}^\dagger   P_L \ell_\alpha\,,
\end{align}
where $\phi$ and $\ell$ are the Higgs and lepton doublets and $\tilde\phi
\equiv i\sigma_2\phi^*$. We work in the mass eigenbasis of the Majorana neutrinos, 
$ M=\mbox{diag}(M_i)$,
which are \textit{CP} eigenstates with parities $\eta_i=\pm 1$, $N_i^c = \eta_i N_i$.
The Majorana condition implies in particular that $\bar N$ and $N$ are not 
independent but related by $\bar N_i =  \eta_i N_i^T C$  where $C = i\gamma_2\gamma_0$ 
is the charge-conjugation matrix. This minimal extension of the SM can
explain naturally the observed small mass scale of the active neutrinos
via the famous see-saw mechanism. Furthermore, \CP--violating 
decays are responsible for the generation of a lepton asymmetry, i.e.~for 
leptogenesis, which is converted into the observed baryon asymmetry by 
the sphaleron processes \cite{Kuzmin:1985mm}.

The \textit{CP}-violating parameter receives contributions 
from the vertex \cite{Fukugita:1986hr} and self-energy
\cite{Flanz:1994yx,Covi:1996wh,Pilaftsis:1997jf,Pilaftsis:2003gt} diagrams. If
the masses of the heavy neutrinos are strongly hierarchical, then the two
contributions are comparable. On the other hand, if the mass spectrum is quasidegenerate
then the self-energy contribution is resonantly enhanced and becomes considerably
larger than the one from the vertex diagram. It has been argued that the resonant enhancement of the
\CP-violating parameter could be compatible with a scale of leptogenesis of the order of
$\sim 1$ TeV \cite{Pilaftsis:2003gt,Pilaftsis:2005rv}.  The lightness of the 
right-handed neutrino masses can in principle induce a detectable non-unitarity of the mixing matrix of active neutrinos \cite{Antusch:2009gn,Antusch:2009pm} which is 
very interesting from the theoretical and experimental point of view. In addition, 
a model with right-handed neutrinos lighter than $10^7$ GeV does not suffer from the hierarchy problem \cite{Boyarsky:2009ix}.

The  self-energy contribution to the \CP-violating parameter has been 
extensively investigated in the literature. One of the first estimates was 
made in \cite{Flanz:1996fb} using the effective Hamiltonian approach known 
from kaon physics. For a quasidegenerate mass spectrum of the heavy 
neutrinos, $M_2\simeq M_1$, the obtained result for the \CP-violating parameters  
can be represented in the form: 
\begin{align}
\label{EpsilonPaschos}
\epsilon_i = \frac{\Im[(\yu^\dag\yu)_{ij}^2]}{8\pi (\yu^\dag\yu)_{ii}}
\,R\,,\quad 
R\equiv\frac{M_i M_j(M^2_i-M^2_j)}{(M^2_i-M^2_j)^2+A^2}\,,
\end{align}
where $i,j=1,2$ are flavor indices, and the resonant enhancement is described by the function
$R$. For $A=0$, the function $R$ would diverge in the degenerate limit. Therefore $A$ can be
considered as a ``regulator'', that arises from the fact that the Majorana neutrinos are
not strictly on-shell because of their finite lifetime. In Ref.~\cite{Flanz:1996fb} it was found
\begin{align}
A={\textstyle \frac1{32\pi}}\text{Re}(\yu^\dag\yu)_{12} (M_1+M_2)^2\,.
\end{align} 
The regulator is relevant for determining the maximal possible resonant enhancement, which
occurs for $M_2^2-M_1^2=\pm A$, and is given by
\begin{align}
\label{Rmax}
R_{\it max} = \frac{M_1M_2}{2|A|} \;.
\end{align}
Note that \eqref{EpsilonPaschos} is proportional to the product 
$M_1M_2\,(M_2^2-M_1^2)\,\Im(\yu^\dag\yu)_{12}^2$ which, similarly to the 
Jarlskog determinant, is an invariant (under field rotations) quantity 
characterizing \CP-violation in the leptonic sector. In particular, if 
the heavy neutrino masses are equal, i.e. if $M_1^2=M_2^2$, then the 
\CP-violating phases in the Yukawa couplings can be rotated away and 
the \CP-violating parameter vanishes.
Extending the formalism developed in \cite{Liu:1993ds} the authors of 
\cite{Covi:1996fm} studied the effects of particle mixing for scalar fields
with arbitrary mass splitting and also found a considerable enhancement of 
the asymmetries when the masses of the mixed states are comparable. 
For decreasing mass splitting the maximal enhancement is achieved for 
$M_2^2-M_1^2\simeq \Gamma^2_{11}+\Gamma^2_{22}$, where $\Gamma_{ij}$ 
denotes the absorptive part of the one-loop self-energy diagram. 

An approach based on an expansion of full resummed propagators around
their poles was developed and applied to the analysis of leptogenesis in the SM in
\cite{Pilaftsis:1997jf,Pilaftsis:1997dr,Pilaftsis:1998pd,Pilaftsis:2003gt}.
A related approach based on the diagonalization of the resummed propagator
was developed in Ref.~\cite{Buchmuller:1997yu}.
The starting point of both approaches is the solution of the Schwinger-Dyson 
equation for the full propagator of the mixing species in vacuum. 
Then the resummed propagators are substituted into  amplitudes of the 
lepton-number violating scattering processes, for which only stable
particles appear in the asymptotic in and out states.
The arising expressions are then interpreted in 
terms of effective decay widths of the Majorana neutrino species
and \CP-violating parameters are extracted.
Within the former approach, the decay rates are obtained by extracting
the on-shell contributions of the Majorana neutrino propagator
to the scattering processes. The resulting \CP-violating parameter has the form 
of Eq.~\eqref{EpsilonPaschos} but with a ``regulator'' given by 
\begin{align}
\label{EpsilonPilaftsis}
A=M_i \Gamma_j \;.
\end{align}
The result obtained for the \CP-violating parameter based on the diagonalization
approach~\cite{Buchmuller:1997yu} also has the form of Eq.~\eqref{EpsilonPaschos},
but with 
\begin{align}
\label{EpsilonAnisimov}
A=M_i\Gamma_i - M_j\Gamma_j\,.
\end{align}
As has been noted in Ref.~\cite{Anisimov:2005hr}, the difference can be traced back
to the way how the \CP-violating decay rates are extracted from the scattering
amplitudes. Furthermore, it has been argued that both the pole mass expansion
and diagonalization approach can be reconciled, and yield a regulator
of the form~\eqref{EpsilonAnisimov}. In general there exist two potential
sources that can contribute to the regulator, arising from the finite width of the two
Majorana neutrino species required for a \CP-violating contribution to the
decay rate. Therefore it seems plausible that the regulator depends on the widths
of both species. As is evident from Eq.~\eqref{Rmax}, the latter result predicts a larger value 
of the \CP-violating parameter, especially if $\Gamma_i\sim \Gamma_j$. 
A comparison of the various approaches to the 
calculation of the \CP-violating parameter in the resonant regime 
can also be found in \cite{Rangarajan:1999kt}.  

\medskip

In all of the above works the \CP-violating parameter is evaluated in vacuum and then used to calculate the asymmetry generated by the decays of the heavy neutrinos. In the hot and dense plasma thermal corrections to the
decay and scattering rates can play an important role. An analysis of leptogenesis within thermal 
quantum field theory was performed in \cite{Covi:1997dr,Giudice:2003jh}. The influence of thermal corrections
on the dispersion relations was studied in \cite{Kiessig:2010pr,Kiessig:2011fw,Kiessig:2011ga}.
The production rate of right-handed neutrinos in a thermal plasma of SM particles has been studied in \cite{Besak:2010fb,Anisimov:2010gy} 
and in \cite{Salvio:2011sf,Laine:2011pq} for temperatures that are large or small compared to their mass, respectively.

The resonant enhancement of the \CP-violating parameter  becomes maximal
when the mass difference of the neutrinos is of the order of their width. Expressed
in a quantum field theoretical language, this means that the spectral functions of
the neutrinos cannot be described by two distinct peaks, but rather by two overlapping
resonances. In addition, this implies that it is not sufficient to describe the
deviation from equilibrium in terms of semi-classical distribution functions 
$f_{N_i}(t)$ for the two neutrinos $i=1,2$. Instead, it is important to take a 
full two-by-two matrix into account, that can describe coherent transitions between 
the neutrino species in terms of cross-correlations. Within a purely quantum field 
theoretical treatment, a natural object to consider are the two-point functions 
$\langle \bar N_i(x) N_j(y) \rangle$, which possess a matrix structure in flavor 
space. Their time-evolution can be described by the so-called closed-time-path (CTP) or Schwinger-Keldysh formalism. Concretely, self-consistent equations  of motion 
for the non-equilibrium time-evolution can be obtained by formulating the 
Schwinger-Dyson equation in real time on the CTP. The resulting equations are 
known as Kadanoff-Baym equations.
One of the first steps towards the analysis of leptogenesis within the 
formalism of non-equilibrium QFT has been performed in \cite{Buchmuller:2000nd}
where the authors solved the system of Kadanoff-Baym equations 
in the Minkowski space-time using a perturbative expansion. In \cite{Anisimov:2008dz} 
the approach to equilibrium for a scalar field which is coupled to 
a large thermal bath was studied.  A quantum mechanical calculation of the asymmetry 
generation for a hierarchical mass spectrum of the heavy neutrinos based 
on Kadanoff-Baym equations was presented by the same group in 
\cite{Anisimov:2010aq,Anisimov:2010dk}. We also refer to Refs.~\cite{DeSimone:2007rw,DeSimone:2007pa}
for previous work on a quantum treatment of leptogenesis.
A somewhat different approach was pursued
in \cite{Garny:2009rv,Garny:2009qn,Garny:2010nj,Garny:2010nz} as well as in 
\cite{Beneke:2010wd,Beneke:2010dz,Garbrecht:2010sz}. In these works the 
initial system of Kadanoff-Baym equations was used to derive Boltzmann-like 
equations for the quasiparticles and calculate their effective in-medium properties
like masses, decay widths and \CP-violating parameters. The vertex and 
self-energy contributions to the \CP-violating parameter in a toy model with 
two real scalar fields and one complex scalar field were analyzed in 
\cite{Garny:2009rv} and \cite{Garny:2009qn}. 
A similar analysis for leptogenesis in the Standard Model
was performed in \cite{Beneke:2010wd,Garbrecht:2010sz}. 
Flavor oscillations in the leptonic sector and their impact on the 
generation of an asymmetry were analyzed within nonequilibrium quantum field theory in 
\cite{Beneke:2010dz}. 
Oscillations of the \textit{right-handed} neutrinos play a crucial role in the 
scenario of leptogenesis via oscillations \cite{Akhmedov:1998qx,Asaka:2005pn,
Canetti:2010aw}. The latter is realized, in particular, in the Neutrino
Minimal Standard Model ($\nu$MSM) \cite{Boyarsky:2009ix}. The calculation of 
the asymmetry generated within this scenario is based on so-called 
density matrix equations \cite{Sigl:1992fn}. Another method using the perturbative 
short-time solution of the von Neumann equation and intended to check 
the consistency of the density matrix equations was presented in 
\cite{Gagnon:2010kt}.

\medskip
 
Summarizing the above, the analysis of leptogenesis in the maximally resonant 
regime poses two problems, which, despite the recent progress, are not 
yet fully resolved. The first one is the enhancement of the \CP-violating 
parameter. The existing vacuum calculations lead to rather different results. 
Furthermore, they neglect the medium effects
as well as modification of the quasiparticle spectra in the hot and 
dense plasma of the early Universe. It is also a priori unclear whether the very 
notion of quasiparticles is applicable in the maximal resonant regime. 
The second problem is the oscillation of the heavy neutrinos which 
becomes important when the mass difference is sufficiently 
small. This effect is completely neglected in the standard scenario 
of resonant thermal leptogenesis which, in the maximal resonant regime,
``merges'' with the scenario of leptogenesis via oscillations. 

In this work we study the 
generation of a lepton asymmetry from first principles 
using an explicit solution of the Kadanoff-Baym equations for mixing 
Majorana neutrinos with an arbitrary mass difference. Following Refs.~\cite{Anisimov:2010aq,Anisimov:2010dk},
we will treat lepton and Higgs fields as a thermal bath, neglecting the backreaction. This
simplification has the advantage that it is possible to obtain analytical results
that can be directly compared to the Boltzmann approach.

In section \ref{sec:ctp}, we introduce the framework of our computation, explain our notations, 
 define the quantities that are necessary for
a quantum-mechanical approach to leptogenesis, and derive the relevant Kadanoff-Baym equations. 
In section \ref{sec:EoMAsymmetry} we derive an equation of motion for the lepton asymmetry, and discuss the
generation of an asymmetry from the interaction of the Majorana neutrinos with
a thermal bath. An analytical solution for the Kadanoff-Baym equations and the lepton
asymmetry is derived in section \ref{sec:BW} using a Breit-Wigner approximation.
In section \ref{sec:result} we present our results obtained in the Kadanoff-Baym approach
and compare it to the corresponding Boltzmann approach. In particular, we discuss the differences that arise with
respect to the resonant enhancement of the lepton asymmetry. We also compare the solutions of the
Kadanoff-Baym equations for various approximations. Finally, we conclude
in section \ref{sec:conclusion}. Details of the derivation of the analytical solution can be
found in the Appendix.

\section{Closed-time-path formalism}\label{sec:ctp}

Leptogenesis is a non-equilibrium process. Within quantum field theory,
such processes can be conveniently described based on the closed-time-path
or Schwinger-Keldysh formalism. In particular, we are interested in the non-equilibrium
time-evolution of the expectation value of physical observables, like
the baryon or lepton asymmetry. Consider for example the lepton current
operator,
\begin{equation}
  J^\mu_L(x) = \sum_\La \bar \lepton_\La(x) \gamma^\mu \lepton_\La(x) \;.
\end{equation}
We are interested in the time-evolution of its expectation value
with respect to the physical state of the system,
\begin{equation}
 j^\mu_L(x) = \langle J^\mu_L(x) \rangle = \left\langle \sum_\La \bar \lepton_\La(x) \gamma^\mu \lepton_\La(x) \right\rangle \;.
\end{equation}
The lepton asymmetry per volume $V$ is then given by
\begin{equation}
  n_L(t) =  \frac{1}{V} \int_V d^3x \, j^0_L(t,\bm{x}) \;.
\end{equation}
It is the main purpose of this work to compute the time-evolution of
the lepton asymmetry that is generated by the out of equilibrium
decays of quasi-degenerate Majorana neutrino species within the
closed time path formalism. In the following, we will introduce the
quantities and equations necessary for this computation.

\subsection{Propagators}

Within quantum field theory a system out of equilibrium can be characterized
completely by its full set of $N$-point correlation functions. Since we
are interested in a temperature regime for which the electroweak symmetry
is restored, the lowest-order correlations are the two-point functions
\begin{eqnarray}
  \pH_{ab}(x,y) & = & \langle T_\CTP \higgs^a(x) \higgs^{*b}(y) \rangle \,, \nonumber \\
  \pL^{\alpha\beta}_{ab}(x,y) & = & \langle T_\CTP \lepton_{\alpha}^a(x) \bar \lepton_{\beta}^b(y) \rangle \,, \\
  \pN^{ij}(x,y) & = & \langle T_\CTP \majneutrino_{i}(x) \bar\majneutrino_{j}(y) \rangle \;.\nonumber
\end{eqnarray}
Here $a,b=1,2$ are $SU(2)$ indices, and $\alpha,\beta=1,2,3$ as well as $i,j=1,2,3$ are flavor indices
of the leptons and Majorana neutrinos respectively.
We use matrix notation for the Lorentz-spinor indices. $T_\CTP$ denotes time-ordering with respect to the
closed time contour, and features the usual minus sign when interchanging fermionic operators.
In an $SU(2)$-symmetric state, which we consider in the following, one can write
\begin{eqnarray}\label{SU2SymmProp}
 \pH_{ab}(x,y) & = & \delta_{ab}\pH(x,y) \,, \nonumber \\
  \pL^{\alpha\beta}_{ab}(x,y) & = & \delta_{ab} \pL^{\alpha\beta}(x,y) \,.
\end{eqnarray}
Note that the lepton and Majorana neutrino propagators are matrices in flavor space, i.e. they encode cross-correlations which are important for coherence effects. Throughout this work we assume that the third Majorana neutrino is very heavy, and therefore consider the two-by-two matrix in flavor space corresponding to $\majneutrino_{1,2}$. We will frequently use matrix notation for the flavor indices.
If the two-point functions are known, the lepton current can be easily calculated,
\begin{equation}
   j_L^\mu(x) = - \sum_\alpha \tr \left[\gamma_\mu \pL^{\alpha\alpha}(x,x) \right]\;.
\end{equation}
Thus, the main task will be to obtain non-equilibrium evolution equations
for the two-point functions, so-called Kadanoff-Baym equations, and to
extract the relevant processes which describe the generation of the
lepton asymmetry.

By denoting the time arguments on the chronological branch of the closed-time
contour by $x^+$ and the one on the anti-chronological branch by $x^-$,
one can obtain four propagators with usual real time arguments,
\begin{equation}
 D_{++}(x,y) = D(x^+,y^+), \quad D_{+-}(x,y) = D(x^+,y^-), \atop
 D_{-+}(x,y) = D(x^-,y^+), \quad D_{--}(x,y) = D(x^-,y^-) \;,
\end{equation}
where $D(x,y)$ stands for any two-point function.
The four functions are often combined into a two-by-two matrix structure
(in addition to the flavor and Dirac indices).
However, it is easy to see that only two of them are independent.
In this work, we prefer to use the more compact notation of~\cite{Danielewicz:1982kk},
and decompose the two-point function into \emph{spectral function} $D_\rho$
and \emph{statistical propagator} $D_F$,
\begin{equation}
  D(x,y) = D_F(x,y) -\frac{i}{2}\sgn_\CTP(x^0-y^0)D_\rho(x,y) \;.
\end{equation}
The signum function is either $+1$ or $-1$ depending on whether
$x^0$ or $y^0$ occur `later' on the closed time contour $\CTP$. From this parameterization
it is evident that in fact only two independent two-point functions exist.
All other two-point functions can be expressed as linear combinations.
For example, one can introduce the Wightman functions
$D_\gtrless(x,y)  =  D_F(x,y) \mp \frac{i}{2}D_\rho(x,y)$.
In terms of the matrix notation, these would correspond to $D_>=D_{-+}$
and $D_<=D_{+-}$. Finally, it is useful to introduce \emph{retarded}
and \emph{advanced} functions,
\begin{eqnarray}
 D_R(x,y) & = & \Theta(x^0-y^0)D_\rho(x,y) \,, \\
 D_A(x,y) & = & - \Theta(y^0-x^0)D_\rho(x,y) \,.
\end{eqnarray}

A useful relation for the Majorana neutrino propagator, which results from the relation $N_i^c=\eta_i N_i$, is
\begin{equation}
  \pN^{ij}(x,y) =  \eta_i\eta_j C \pN^{ji}(y,x)^T C^{-1} \;,
\end{equation}
implying that
\begin{eqnarray*}
  \pN^{ij}_{F,R/A}(x,y) & = &   \eta_i\eta_j C \pN^{ji}_{F,A/R}(y,x)^T C^{-1}\;, \\
  \pN^{ij}_{\rho}(x,y) & = & {} - \eta_i\eta_j C \pN^{ji}_{\rho}(y,x)^T C^{-1}\;.
\end{eqnarray*}

\subsubsection*{\CP-conjugate propagators}

For leptogenesis, it is useful to consider separately \CP-odd
and \CP-even effects. In particular, the generation of the
lepton asymmetry is a \CP-odd effect. In order to discriminate between
\CP-odd and \CP-even, one can perform a \textit{CP} transformation of
the system, and determine the transformation properties of the
two-point functions.

The Higgs field transforms like $\phi(x) \to \phi^*(\bar x)$ under \textit{CP}, where $\bar x\equiv(x^0,-\vec{x})$.
Thus the propagator gets transformed in the following way:
\begin{equation}\nonumber
  \pH_{ab}(x,y) = \langle T_\CTP \phi_a(x) \phi_{b}^*(y) \rangle \to \langle T_\CTP \phi_{a}^*(\bar x) \phi_{b}(\bar y) \rangle = \pH_{ba}(\bar y,\bar x) \;.
\end{equation}
Motivated by this observation, we define the \emph{CP-conjugate propagator} as
\begin{equation}
  \pHcp_{ab}(x,y) \equiv \pH_{ba}(\bar y,\bar x) \;.
\end{equation}
From this, one can deduce the various components introduced above,
\begin{eqnarray*}
  \pHcp_{ab\,F,R/A}(x,y) & = &  \pH_{ba\,F,A/R}(\bar y,\bar x) \;, \\
  \pHcp_{ab\,\rho}(x,y) & = &  - \pH_{ba\,\rho}(\bar y,\bar x) \;.
\end{eqnarray*}
It is important to note that for a system in a \emph{CP-symmetric state}
the propagator and the \textit{CP}-conjugate propagator are equal, $\pHcp_{ab}(x,y)=\pH_{ab}(x,y)$.

\medskip

The lepton field transforms like $\lepton(x) \to CP \bar \lepton^T(\bar x)$ and $\bar \lepton(x) \to \lepton^T(\bar x)PC = - \lepton^T(\bar x)(CP)^{-1}$, where $P = \gamma_0$. Accordingly, the propagator gets transformed like
\begin{equation}\nonumber
  \pL^{\La\Lb}_{\Da\Db\,ab}(x,y) \to (CP)_{\Da\Dc} \pL_{\Dd\Dc\,ba}^{\Lb\La}(\bar y,\bar x) [(CP)^{-1}]_{\Dd\Db} \;,
\end{equation}
where we have displayed Dirac indices for clarity.
The \textit{CP}-conjugate propagator is then given by
\begin{equation}
  \pLcp^{\La\Lb}_{ab}(x,y) \equiv CP \pL^{\Lb\La}_{ba}(\bar y,\bar x)^T (CP)^{-1} \;.
\end{equation}
Here we have again used a matrix-notation for the Lorentz-spinor indices, and the transpose
on the right-hand side refers to these indices. For the components, one finds
similarly as above
\begin{eqnarray*}
  {\pLcp}^{\La\Lb}_{ab\, \gtrless,F,R/A}(x,y) & = &  CP \pL^{\Lb\La}_{ba\, \lessgtr,F,A/R}(\bar y,\bar x)^T (CP)^{-1} \;, \\
  {\pLcp}^{\La\Lb}_{ab\, \rho}(x,y) & = &  - CP \pL^{\Lb\La}_{ba\, \rho}(\bar y,\bar x)^T (CP)^{-1} \;.
\end{eqnarray*}
For a system in a \emph{CP-symmetric state}, one has $\pLcp^{\La\Lb}_{ab}(x,y)  = \pL^{\La\Lb}_{ab}(x,y)$.

\medskip

The \textit{CP}-conjugate propagator of the Majorana neutrinos is given by
\begin{equation}
  \pNcp^{ij}(x,y) = CP \pN^{ji}(\bar y,\bar x)^T (CP)^{-1} \;.
\end{equation}
For the components, the relations are also analogous to the leptons.

\subsection{Kadanoff-Baym equations}

The non-equilibrium equation of motion for the two-point functions
can be derived by a variational principle from the so-called two-particle-irreducible (2PI)
effective action $\Gamma[\pH,\pL,\pN]$. Technically, it is a Legendre transform with respect to
the generating functional in the presence of bi-local sources. All the information about the
quantum system is encoded in the effective action.

\begin{figure}[t]
  \begin{center}
    \feyn{0mm}{0.16}{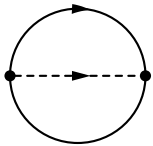}
  \end{center}
  \caption{\label{fig:EffectiveAction}
  Two-loop diagram contributing to the 2PI effective action.}
\end{figure}

The equation of motion for the two-point functions
follows from the stationarity conditions,
\begin{equation}
  \frac{\delta\Gamma}{\delta \pH} = 0 , \quad \frac{\delta\Gamma}{\delta \pL} = 0 , \quad \frac{\delta\Gamma}{\delta \pN} = 0 \;.
\end{equation}
In general, the 2PI effective action can be parameterized in the form
\begin{equation}
  \Gamma[\pH,\pL,\pN] = {\cal S} + \Gamma^{1L}[\pH,\pL,\pN] + \Gamma_2[\pH,\pL,\pN] \,,
\end{equation}
where the classical action ${\cal S}=\int_\CTP d^4x {\cal L}$ is the tree-level
contribution, $\Gamma^{1L}$ the one-loop contribution, and $\Gamma_2$ contains
diagrams with at least two loops. More precisely, $\Gamma_2[\pH,\pL,\pN]$ is equal
to the sum of all 2PI (`skeleton') Feynman diagrams which have no external legs.
Furthermore, the internal lines represent the \emph{full} propagators $\pH,\pL$, and $\pN$.
The two-loop contributions are shown in figure \ref{fig:EffectiveAction}.

\begin{figure}[t]
  \begin{center}
    \feyn{0mm}{0.8}{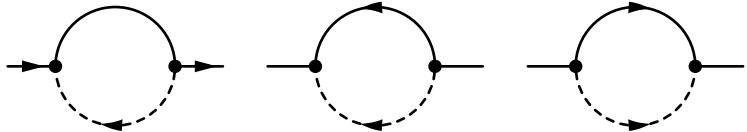}
  \end{center}
  \caption{\label{fig:SelfEnergy}
  One-loop diagrams contributing to the lepton (left) and Majorana neutrino (middle and right) self-energies.}
\end{figure}

The resulting equations have the form of Schwinger-Dyson equations,
evaluated on the closed-time path. For the leptons and the Majorana neutrinos, they
read (suppressing flavor, SU(2) and Lorentz-spinor indices for the moment)
\begin{subequations}
\begin{align}
  \frac{\delta\Gamma}{\delta \pL} = 0 \quad \Leftrightarrow \quad \pL^{-1}(x,y) & =  \pL_0^{-1}(x,y) - \seL(x,y) \;,  \\
  \frac{\delta\Gamma}{\delta \pN} = 0 \quad \Leftrightarrow \quad \pN^{-1}(x,y) & =  \pN_0^{-1}(x,y) - \seN(x,y) \;. 
\end{align}
\end{subequations}
Here $\pL^{-1}(x,y)$ is the inverse of the \emph{full} lepton
two-point function (i.e. the inverse of the full propagator), $\pL_0^{-1}(x,y)$
is the inverse of the \emph{free} lepton propagator\footnote{Strictly speaking, in Dirac notation the inverse exists
only when restricting the spinors to the left-handed sub-space. Nevertheless it is possible to work in Dirac notation
when properly inserting left-handed projectors $P_L$, so that effectively only the left-handed parts contribute.},
\begin{equation}
  i\pL_0^{-1}(x,y) = i\slashed{\partial}_x \delta_{ab}\delta^{\La\Lb} \delta_\CTP(x-y) P_L \,,
\end{equation}
and $\seL(x,y)$ is the self-energy given by the functional derivative of the
effective action $\Gamma_2[\pH,\pL,\pN]$,
\begin{equation}
  \seL^{\La\Lb}_{\Da\Db\,ab}(x,y) = \frac{i\delta\Gamma_2}{\delta \pL^{\Lb\La}_{\Db\Da\,ba}(y,x)} \,.
\end{equation}
Graphically, this corresponds to opening one lepton line of the diagrams contributing
to $\Gamma_2$. Thus, for example the two-loop contribution to $\Gamma_2$ yields the
one-loop self-energy. The one-loop contribution to the lepton self-energy is
shown in figure \ref{fig:SelfEnergy}. We stress again that the
internal lines of the diagrams represent the \emph{full} propagators. This means
that the equation of motion is a \emph{self-consistent} Schwinger-Dyson equation.

Similarly, the free inverse propagator for the Majorana neutrinos reads
\begin{equation}
  i\pN_0^{-1}(x,y) = \left(i\slashed{\partial}_x - M_i \right) \delta^{ij} \delta_\CTP(x-y) \,,
\end{equation}
Later on, it will be necessary to cancel divergent contributions to the self-energy of the
Majorana neutrinos. Therefore, we introduce suitable mass and field counterterms. Their most
general form can be obtained by applying a field rescaling $N_i \to (Z_{ij}P_L + \eta_i Z_{ij}^* \eta_j P_R) N_j$
compatible with the Majorana conditions and using the most general form of the mass matrix for the
mass counterterms, $\delta  M_{ij} =\eta_i\delta  M_{ji}\eta_j$. The resulting counterterms are
given by \cite{Pilaftsis:2003gt,Kniehl:1996bd}
\begin{eqnarray}
\label{Counterterm}
\delta{\cal L}
& = & {\frac12}\bar\majneutrino\bigg\{ {\frac{i}{2}}
\big( \delta   Z^\dagger + \delta  Z\big) P_R \slashed{\partial} + {\frac{i}2} \big(\eta\delta Z^T\eta + \eta\delta  Z^*\eta\big) P_L  \slashed{\partial} - {\frac12}\big( \delta  Z^\dagger  M+2\delta M^\dag\nonumber\\
&& {} + M\eta \delta  Z^*\eta\big)P_R - {\frac12}\big(\eta\delta  Z^T\eta  M+2\delta  M+ M \delta  Z\big)P_L\bigg\}\majneutrino\,,
\end{eqnarray}
The (renormalized) self-energy has two contributions,
\begin{equation}
  \seN^{ij}_{\Da\Db}(x,y) = \frac{i\delta\Gamma_2}{\delta \pN^{ji}_{\Db\Da}(y,x)} + \delta\seN^{ij}_{\Da\Db}(x,y)\,.
\end{equation}
The latter arises from the counterterms (\ref{Counterterm}),
\begin{equation}
  \delta\seN^{ij}(x,y) = -i\delta\seN^{ij}(x)\delta_\CTP(x-y) \,,
\end{equation}
\begin{align}
\delta{\seN}(x)
& =  - \Big\{{\textstyle\frac{i}2}
( \delta   Z^\dagger + \delta Z) P_R \slashed{\partial}_x +{\textstyle\frac{i}2} (\eta\delta  Z^T\hat\eta + \eta\delta  Z^*\eta) P_L \slashed{\partial}_x-{\textstyle\frac12}( \delta  Z^\dagger  M\nonumber\\
&+2\delta M^\dag + M\eta \delta  Z^*\eta)P_R -{\textstyle\frac12}(\eta\delta  Z^T\eta  M+2\delta  M+ M \delta  Z)P_L\Big\} \,,
\end{align}
The Majorana self-energy inherits the relation
\begin{equation}
  \seN^{ij}(x,y) = \eta_i\eta_j C \seN^{ji}(y,x)^T C^{-1} \;,
\end{equation}
of the Majorana propagator. Note that also the counterterm-part fulfills this relation.

The equation of motion can be brought in a more explicit form by convoluting it
with the full propagator, and inserting the decomposition into statistical and
spectral components. Then one obtains a system of two coupled integro-differential
equations, so-called Kadanoff-Baym equations,
\begin{subequations}
\begin{eqnarray}
\label{KBEL}
  i \slashed{\partial}_x \pL^{\La\Lb}_{F}(x,y)    & = & \int_0^{x^0} d^4z \, \seL^{\La\Lc}_{\rho}(x,z) \pL^{\Lc\Lb}_{F}(z,y) \nonumber\\
                                           &   & {} - \int_0^{y^0} d^4z \, \seL^{\La\Lc}_{F}(x,z) \pL^{\Lc\Lb}_{\rho}(z,y) \;, \\
  i \slashed{\partial}_x \pL^{\La\Lb}_{\rho}(x,y) & = & \int_{y^0}^{x^0} d^4z \, \seL^{\La\Lc}_{\rho}(x,z) \pL^{\Lc\Lb}_{\rho}(z,y) \;.
\end{eqnarray}
\end{subequations}
The corresponding Kadanoff-Baym equation for the Majorana neutrinos reads (in matrix notation for the flavor indices)
\begin{subequations}
\begin{eqnarray}\label{KBEN}
  (i \slashed{\partial}_x - M - \delta{\seN}(x)) \pN_{F}(x,y)    & = & \int_0^{x^0} d^4z \, \seN_{\rho}(x,z) \pN_{F}(z,y) \nonumber\\
                                           &   & {} - \int_0^{y^0} d^4z \, \seN_{F}(x,z) \pN_{\rho}(z,y) \;, \\
  (i \slashed{\partial}_x - M - \delta{\seN}(x)) \pN_{\rho}(x,y) & = & \int_{y^0}^{x^0} d^4z \, \seN_{\rho}(x,z) \pN_{\rho}(z,y) \;.\label{KBENspec}
\end{eqnarray}
\end{subequations}
These equations contain `memory integrals' on the right-hand side, which integrate over the history
of the system. Thus, the change of the two-point function at a given instant of time depends on the
entire history of the system, i.e. the equations are non-Markovian. Furthermore, the equations
explicitly depend on the two time-arguments $x^0$ and $y^0$ characterizing the two-point function.
For a system in equilibrium, they would only depend on the difference $s^0 = x^0-y^0$. Out of equilibrium,
the system also depends on the average time $X^0 = (x^0+y^0)/2$. For this reason, the equations in the form given above
are sometimes referred to as \emph{two-time Kadanoff-Baym equations}. We stress that these equations
are valid also far from equilibrium. For systems which are close to equilibrium, one expects that
the two-point functions depend only weakly on $X^0$. In this case, it is customary to perform a gradient
expansion, which yields Markovian quantum kinetic equations. In addition, one may perform a Fourier
transformation with respect to $s^0$ (Wigner transformation). Then, a very common further simplification
consists of making a particular ansatz for the frequency-dependence of the two-point functions, including
a Breit-Wigner ansatz or a quasi-particle ansatz. The latter is known to lead to Boltzmann-type equations.

On the other hand, it is also possible to follow a different strategy, namely to retain the non-Markovian,
two-time structure of the equations. If, in addition, the self-energies are computed using the complete
propagators, the resulting equations are non-linear integro-differential equations. Equations of this
type have been solved numerically with increasing complexity over the past years, and it has been shown that they can describe the
process of kinetic and chemical quantum thermalization for systems that are initially far from
equilibrium, see e.g. \cite{Berges:2000ur,Berges:2001fi,Aarts:2001yn,Berges:2002wr,Juchem:2003bi,Arrizabalaga:2005tf,Lindner:2005kv}. A significant simplification can be achieved when treating the self-energies as time-translation invariant functions, which reduces the Kadanoff-Baym equations to linear integro-differential equations. Physically, this can occur when the quantum field that is out of equilibrium is weakly interacting with a thermal bath, and the interaction can be described by self-energies that solely depend on propagators associated to particle species belonging to the thermal bath. The approximation
of time-translation invariant self-energies then corresponds to neglecting the backreaction of the non-equilibrium quantum field on the thermal bath (more precisely, to neglect the possible deviation from equilibrium of the self-energies that is
caused by the backreaction). For that case, approximate analytical solutions of the Kadanoff-Baym equations within
the two-time representation have been derived in \cite{Anisimov:2008dz}. Moreover, these solutions turn out to be well-suited for describing
the evolution of the Majorana neutrino propagator, and have been used to study leptogenesis in the hierarchical limit \cite{Anisimov:2010aq,Anisimov:2010dk}. The reason is twofold: First, the corresponding self-energies depend on the lepton and Higgs propagators, and the latter belong to the thermal bath of $\mathcal{O}(100)$ SM particles that are kept in equilibrium due to gauge interactions. Second, the Majorana neutrino itself interacts weakly with the SM via the Yukawa couplings. In the present work, we will follow the same strategy to obtain analytical
solutions of the Kadanoff-Baym equations for the resonant case.

Although the two-time Kadanoff-Baym equations shown above are valid far from equilibrium and
do not require any a priori assumptions of the spectral properties, they do have a restriction
which is not so commonly emphasized. Namely, the equations in the above form implicitly assume
that the initial state of the system, corresponding to $x^0=y^0=0$, is a \emph{Gaussian} initial
state. This means that the connected part of all higher correlation functions (three-, four-point
correlations and so on) vanish at the initial time. Physically, this means that the interactions
are `switched on' just at the time $t=0$. However, it is possible to generalize the Kadanoff-Baym equations
such that they can also describe systems with more realistic initial conditions, if necessary \cite{Garny:2009ni}. Below, we will in fact include higher correlations in an effective way
by taking the evolution of the system before the `initial' time into account.

\subsection{Self-energies}

The self-energies, which enter into the Kadanoff-Baym equations on the right-hand side,
contain all the information about the interactions of the system.
As discussed above, the one-loop self-energy is obtained from the
two-loop contribution to the 2PI effective action (see figure \ref{fig:EffectiveAction}),
\begin{eqnarray}
	\Gamma_2^{2L}[\pH,\pL,\pN] & = & \sum_{ij\La\Lb} \yu_{i\La}^\dag\yu_{\Lb j} \int_\CTP d^4x \int_\CTP d^4y \, \tr \left[ \pN^{ji}(y,x) P_L \pL^{\La\Lb}_{ac}(x,y) P_R \right] \nonumber \\
	                          &   & {} \times \epsilon^{ba} \epsilon^{cd} \pH_{bd}(x,y) \;.
\end{eqnarray}
The one-loop lepton self-energy obtained from differentiating $\Gamma_2^{2L}$ with respect to
the full lepton propagator $\pL(y,x)$ reads (see first diagram in figure \ref{fig:SelfEnergy}),
\begin{eqnarray}
	\seL^{\alpha\beta}_{ab}(x,y)_{1L} & = & \yu_{\La i }\yu_{j\Lb}^\dag  P_R \pN^{ij}(x,y) P_L  \epsilon^{cb} \epsilon^{ad} \pH_{cd}(y,x) \;,
\end{eqnarray}
where summation over $i,j$ is implied.
For a SU(2)-symmetric state, one can insert that $\pH_{ab}=\delta_{ab}\pH$ and $\pL^{\La\Lb}_{ab}=\delta_{ab} \pL^{\La\Lb}$.
Then the self-energy is itself proportional to $\delta_{ab}$, and one can write $\seL^{\La\Lb}_{ab}=\delta_{ab}\seL^{\La\Lb}$,
with
\begin{equation}\label{selfEnergyCTPSeL}
	\seL^{\La\Lb}(x,y)_{1L}  =  - \yu_{\La i }\yu_{j\Lb}^\dag  P_R \pN^{ij}(x,y) P_L \pH(y,x) \;.
\end{equation}

The one-loop contribution to the self-energy of the Majorana neutrinos reads
\begin{eqnarray}
	\seN^{ij}(x,y)_{1L} & = & {} \yu_{i\alpha}^\dag\yu_{\beta j} P_L \pL^{\alpha\beta}_{ac}(x,y) P_R  \epsilon^{ba}\epsilon^{cd} \pH_{bd}(x,y)\nonumber\\
	&& \eta_i\eta_j \yu_{i\alpha}^T\yu_{\beta j}^* P P_L \pLcp^{\alpha\beta}_{ac}(\bar x,\bar y) P_R P \epsilon^{ba}\epsilon^{cd}\pHcp_{bd}(\bar x,\bar y)\;.
\end{eqnarray}
Here summation over $\alpha,\beta$ is implied. After contraction of SU(2) indices
\begin{eqnarray}
\label{selfEnergyCTPSeN}
	\seN^{ij}(x,y)_{1L} & = & {} -2 \yu_{i\alpha}^\dag\yu_{\beta j} P_L \pLH^{\alpha\beta}(x,y) P_R \nonumber\\
	&& {} - 2 \eta_i\eta_j\yu_{i\alpha}^T\yu_{\beta j}^* P P_L \pLHcp^{\alpha\beta}(\bar x,\bar y) P_R P \,,
\end{eqnarray}
where we have introduced an abbreviation for the lepton-Higgs loop
\begin{eqnarray}\label{leptonHiggsLoop}
  \pLH^{\La\Lb}(x,y) & \equiv & \pL^{\La\Lb}(x,y)\pH(x,y) \;, \nonumber\\
  \pLHcp^{\La\Lb}(x,y) & \equiv & CP \pLH^{\Lb\La}(\bar y,\bar x)^T (CP)^{-1} = \pLcp^{\La\Lb}(x,y)\pHcp(x,y) \;.
\end{eqnarray}

Similar to the propagators, it is useful to define \textit{CP}-conjugate self-energies,
\begin{eqnarray}
  \seLcp^{\La\Lb}(x,y) & \equiv & CP \seL^{\Lb\La}(\bar y,\bar x)^T (CP)^{-1} \;, \nonumber \\
  \seNcp^{ij}(x,y) & \equiv & CP \seN^{ji}(\bar y,\bar x)^T (CP)^{-1} \;.
\end{eqnarray}
Then one finds for the \textit{CP}-conjugated one-loop lepton self-energy,
\begin{eqnarray}
	\seLcp^{\La\Lb}(x,y)_{1L} & = & - \yu_{\La i }^*\yu_{j\Lb}^T  P_R \pNcp^{ij}(x,y) P_L \pHcp(y,x) \,.
\end{eqnarray}
and for the Majorana neutrino
\begin{eqnarray}
	\seNcp^{ij}(x,y)_{1L} & = & {} -2 \yu_{i\alpha}^T\yu_{\beta j}^* P_L \pLHcp^{\alpha\beta}(x,y) P_R \nonumber\\
	&& {} - 2 \eta_i\eta_j\yu_{i\alpha}^\dag\yu_{\beta j} P P_L \pLH^{\alpha\beta}(\bar x,\bar y) P_R P \,.
\end{eqnarray}

\section{Equation of motion for the lepton asymmetry}\label{sec:EoMAsymmetry}

The most important observable for leptogenesis is the lepton
asymmetry. Its time-evolution is usually computed by subtracting
the lepton and anti-lepton particle densities. However, since
particle number distributions are a classical concept,
we have to replace them by a suitable quantum-mechanical observable.
A suitable starting point is the operator
\begin{equation}
  J_{L\La\Lb}^\mu(x) = \bar \lepton_\La(x) \gamma^\mu \lepton_\Lb(x) \;,
\end{equation}
which corresponds to the Noether-current for the lepton flavor $\La$
for $\La=\Lb$. We further define a time-dependent lepton-number matrix by
\begin{equation}
  {n_L}_{\La\Lb}(t) = \frac{1}{V} \int_V d^3x \langle J_{L\La\Lb}^0(t,\vec{x}) \rangle \;.
\end{equation}
The entries on the diagonal correspond to the lepton asymmetry
in flavor $\La$ at time $t$ per volume $V$.
The total lepton asymmetry per volume summed over all flavors is
\begin{equation}
  n_L(t) = \sum_\La {n_L}_{\La\La}(t) \,,
\end{equation}
which is independent of the choice of the flavor basis. 
The expectation value of the lepton current operator can be related to the lepton propagator,
\begin{equation}
  j_{L\La\Lb}^\mu(x) \equiv \langle J_{L\La\Lb}^\mu(x) \rangle = \langle \bar \lepton_\La(x) \gamma^\mu \lepton_\Lb(x)  \rangle = -\tr\left[\gamma_\mu \pL^{\La\Lb}(x,x) \right] \;.
\end{equation}
In the presence of lepton-number violating processes, its
divergence can be non-zero. In terms of the propagator, it
can be written as
\begin{eqnarray}
  \partial_\mu j_{L\La\Lb}^\mu(x) & = & -\tr\left[\gamma_\mu (\partial_x^\mu + \partial_y^\mu ) \pL^{\La\Lb}(x,y) \right]_{x=y} \nonumber \\
                             & = & -\tr\left[\slashed{\partial}_x^\mu \pL^{\La\Lb}(x,y) + \pL^{\La\Lb}(x,y) \slashed{\partial}_y^\mu \right]_{x=y} \nonumber \\
                             & = & -\int_\CTP d^4z \, \tr\left[ {\pL_{0}^{-1}}^{\La\Lc}(x,z) \pL^{\Lc\Lb}(z,y) - \pL^{\La\Lc}(x,z) {\pL_{0}^{-1}}^{\Lc\Lb}(z,y) \right]_{x=y} \;.
\end{eqnarray}
\begin{align}
  \partial_\mu & j_{L\La\Lb}^\mu(x)  =  -\tr\left[\gamma_\mu (\partial_x^\mu + \partial_y^\mu ) \pL^{\La\Lb}(x,y) \right]_{x=y} \nonumber \\
                             & =  -\tr\left[\slashed{\partial}_x^\mu \pL^{\La\Lb}(x,y) + \pL^{\La\Lb}(x,y) \slashed{\partial}_y^\mu \right]_{x=y} \nonumber \\
                             & =  -\int_\CTP d^4z \, \tr\left[ {\pL_{0}^{-1}}^{\La\Lc}(x,z) \pL^{\Lc\Lb}(z,y) - \pL^{\La\Lc}(x,z) {\pL_{0}^{-1}}^{\Lc\Lb}(z,y) \right]_{x=y} \;.
\end{align}

In the last line we have inserted the free inverse lepton propagator ${\pL_{0}^{-1}}^{\La\Lb}(x,y)=\delta^{\La\Lb}\slashed{\partial}_x \delta_\CTP(x-y) P_L$
and used the chiral symmetry relations $\pL^{\La\Lb}=P_L\pL^{\La\Lb}=\pL^{\La\Lb}P_R$. Now, one can replace the free inverse propagator $\pL_0^{-1}$ 
on the right-hand side by using the Schwinger-Dyson equation $\pL^{-1}=\pL_0^{-1}-\seL$.
The parts containing $\pL^{-1}\pL$ do cancel, and we obtain the following equation of motion
for the lepton current:
\begin{equation}
  \partial_\mu j_{L\La\Lb}^\mu(x)  =  - \int_\CTP d^4z \, \tr\left[ \seL^{\La\Lc}(x,z) \pL^{\Lc\Lb}(z,x) - \pL^{\La\Lc}(x,z) \seL^{\Lc\Lb}(z,x) \right] \;.
\end{equation}
From this, one can immediately derive an equation of motion for the lepton asymmetry
(assuming a spatially homogeneous system for which $\partial_\mu j^\mu=\partial_0 j^0$),
\begin{equation}\label{EoMLeptonAsymmetry1}
  \frac{ d {n_L}_{\La\Lb}(t) }{ dt }  =  - \frac{1}{V} \int d^3x \int_\CTP d^4y \, \tr\left[ \seL^{\La\Lc}(x,y) \pL^{\Lc\Lb}(y,x) - \pL^{\La\Lc}(x,y) \seL^{\Lc\Lb}(y,x) \right] \;,
\end{equation}
where $x=(t,\vec{x})$.
This can be considered as the master equation for the quantum-mechanical
treatment of leptogenesis.
For future reference, we also remark that one can rewrite the equation equivalently as
\begin{equation}\label{EoMLeptonAsymmetry}
  \frac{ d {n_L}_{\La\Lb}(t) }{ dt }  =  - \frac{1}{V} \int d^3x \int_\CTP d^4y \, \tr\left[ \seL^{\La\Lc}(x,y) \pL^{\Lc\Lb}(y,x) - \seLcp^{\Lb\Lc}(x,y) \pLcp^{\Lc\La}(y,x) \right] \;,
\end{equation}
where we have used that $\tr \seLcp^{\Lb\Lc}(\bar x,\bar y) \pLcp^{\Lc\La}(\bar y,\bar x)=\tr \pL^{\La\Lc}(x,y) \seL^{\Lc\Lb}(y,x)$, and substituted
$\vec{x}\to -\vec{x}$ and $\vec{y}\to -\vec{y}$ in the spatial integrals for the second term.

The `master' equation (\ref{EoMLeptonAsymmetry1}) can be brought in a more explicit form
by inserting the decomposition into statistical and spectral components.
In addition, for a spatially homogeneous system the two-point functions depend only on the
difference of the spatial coordinates $\vec{x}-\vec{y}$. Then it is convenient to switch
to Fourier space in the spatial coordinates,
\begin{align}
  \frac{ d {n_L}_{\La\Lb}(t) }{ dt } & =   i \int \frac{d^3p}{(2\pi)^3} \int_0^t dt' \, \nonumber\\
  &\times \tr\Big[ \seL^{\La\Lc}_{\rho\, \vec{p}}(t,t') \pL^{\Lc\Lb}_{F\, \vec{p}}(t',t) - \seL^{\La\Lc}_{F\, \vec{p}}(t,t') \pL^{\Lc\Lb}_{\rho\, \vec{p}}(t',t) \nonumber\\
  &  - \pL^{\La\Lc}_{\rho\, \vec{p}}(t,t') \seL^{\Lc\Lb}_{F\, \vec{p}}(t',t)  + \pL^{\La\Lc}_{F\, \vec{p}}(t,t') \seL^{\Lc\Lb}_{\rho\, \vec{p}}(t',t)\Big] \;.
\end{align}
By integrating the upper equation with respect to $t$, one can easily obtain
the time-evolution of the lepton asymmetry ${n_L}_{\La\Lb}(t)$,
\begin{align}
  {n_L}_{\La\Lb}(t) & =   i \int \frac{d^3p}{(2\pi)^3} \int_0^t dt' \int_0^{t'} dt'' \, \nonumber\\
  &\times \tr\Big[ \seL^{\La\Lc}_{\rho\, \vec{p}}(t',t'') \pL^{\Lc\Lb}_{F\, \vec{p}}(t'',t') - \seL^{\La\Lc}_{F\, \vec{p}}(t',t'') \pL^{\Lc\Lb}_{\rho\, \vec{p}}(t'',t') \nonumber\\
  & - \pL^{\La\Lc}_{\rho\, \vec{p}}(t',t'') \seL^{\Lc\Lb}_{F\, \vec{p}}(t'',t')  + \pL^{\La\Lc}_{F\, \vec{p}}(t',t'') \seL^{\Lc\Lb}_{\rho\, \vec{p}}(t'',t')\Big] \;.
\end{align}

If the lepton propagator is approximately flavor-diagonal, $\pL^{\La\Lb}=\delta^{\La\Lb}\pL$, it is easy to
see that the integrand is symmetric with respect to $t' \leftrightarrow t''$.
Using that $\int_0^t dt' \int_0^{t'} dt'' (f(t',t'')+f(t'',t')) = \int_0^t dt' \int_0^{t} dt'' f(t',t'')$,
it is possible to obtain a simplified equation,
\begin{align}
  {n_L}_{\La\Lb}(t) = i \int \frac{d^3p}{(2\pi)^3} \int_0^t dt' \int_0^t dt'' \, \tr\Big[ & \seL^{\La\Lb}_{\rho\, \vec{p}}(t',t'') \pL_{F\,\vec{p}}(t'',t') \nonumber\\
            {} - & \seL^{\La\Lb}_{F\, \vec{p}}(t',t'') \pL_{\rho\,\vec{p}}(t'',t')\Big] \;.
\end{align}
This equation essentially agrees with Eq.\,(20) of \cite{Anisimov:2010aq}.
However, we find it more convenient to use Eq.\,(\ref{EoMLeptonAsymmetry}) in the following.

\subsection{Generation of an asymmetry}

In the following, we will explicitly compute the time-evolution of the lepton asymmetry
in the full two-time approach using the two-loop truncation of the 2PI effective action.
For simplicity, we will neglect washout effects and
neglect the expansion of the universe for the moment. Furthermore, we will assume
that the generated asymmetry is small, such that one can expand around a \textit{CP}-symmetric state.
For that purpose, we parameterize the two-point functions and their \textit{CP}-conjugates as
\begin{eqnarray}
  D(x,y) & = & D^s(x,y) + \frac12 \delta D(x,y) \,,\\
  \bar D(x,y) & = & D^s(x,y) - \frac12 \delta D(x,y) \,,
\end{eqnarray}
where $D \in \{\pH,\pL\}$. For a \textit{CP}-symmetric state (i.e. with no lepton asymmetry),
$\delta D$ vanishes. Inserting this parameterization into the equation of motion (\ref{EoMLeptonAsymmetry})
for the lepton asymmetry, and expanding in $\delta D$, yields at leading order
\begin{equation}
  \frac{ d {n_L}_{\La\Lb}(t) }{ dt }  =  -\frac{1}{V} \int d^3x \int_\CTP d^4y \, \tr\left[ \left( \seL^{\La\Lb}(x,y)  - \seLcp^{\Lb\La}(x,y) \right)|_s \pL^s(y,x) \right] \;,
\end{equation}
where we have assumed $\pL^{\La\Lb} \approx  \delta^{\La\Lb} \pL$, and the propagators entering the self-energies are to be evaluated with
the symmetric parts of the propagators, $D^s$, as indicated by the subscript. The washout terms would contribute at
next-to leading order in $\delta D$.

\medskip

Now we are ready to insert the explicit one-loop expressions for the lepton self-energy. This yields the following equation for the lepton asymmetry:
\begin{equation}
  \frac{ d {n_L}_{\La\Lb}(t) }{ dt }  =  \frac{\yu_{\La i }\yu_{j\Lb}^\dag}{V} \int d^3x \int_\CTP d^4y \, \tr \Big[   P_R \left( \pN^{ij}(x,y) - \pNcp^{ji}(x,y) \right) P_L \pLH(y,x) \Big] \;.
\end{equation}
For brevity, we have dropped the superscript '$s$' on lepton and Higgs propagators, and used the abbreviation $\pLH(x,y)=\pL(x,y)\pH(x,y)$ for the lepton-Higgs loop.
By performing manipulations similar to the ones above,
\begin{align}\label{AsymmetryWaveContribution}
  {n_L}_{\La\Lb}(t) = -i\yu_{\La i }\yu_{j\Lb}^\dag \int & \frac{d^3p}{(2\pi)^3} \int_0^t dt' \int_0^t dt'' \nonumber\\
  \tr \Big[ &  P_R \left( \pN^{ij}_{\rho\, \vec{p}}(t',t'') - \pNcp^{ji}_{\rho\, \vec{p}}(t',t'') \right) P_L \pLH_{F\,\vec{p}}(t'',t') \nonumber\\
   - & P_R \left( \pN^{ij}_{F\, \vec{p}}(t',t'') - \pNcp^{ji}_{F\, \vec{p}}(t',t'') \right) P_L \pLH_{\rho\,\vec{p}}(t'',t') \Big] \;.
\end{align}
This equation provides a quantum field theoretical expression for the lepton asymmetry that is
generated from the wave or self-energy type contributions. It is appropriate to study the resonant case, where these contributions dominate. As input, it requires the two-point
functions for the lepton, Higgs and Majorana neutrino fields. Since lepton and
Higgs interact much more strongly than the Majorana neutrino, it is often assumed
that they are in thermal equilibrium. Thus, one needs to provide (at least) the
equilibrium expressions for the lepton and Higgs propagators, as well as the
non-equilibrium Majorana propagator. In particular, the information on the \textit{CP} violation is encoded in the off-diagonal components of the Majorana propagator. In the following, we will evaluate the
lepton asymmetry for different choices for these propagators, that correspond to
different approximation levels, and enable a comparison to the conventional Boltzmann result.

\subsection{Asymmetry in a thermal bath}\label{sec:ThermalBath}

In the following, we will derive an expression for the lepton asymmetry
assuming that lepton and Higgs fields can be treated as a thermal
bath, and that the back-reaction can be neglected. This setup has
been considered in Refs.~\cite{Anisimov:2010dk,Anisimov:2010dk} for a hierarchical
neutrino spectrum. Concretely, this means
that we assume that the two-point functions for lepton and Higgs fields are
time-translation invariant,
\begin{eqnarray}
 \pL(x,y) & = & \pL^{th}(x-y) \,, \nonumber\\
 \pH(x,y) & = & \pH^{th}(x-y) \,.
\end{eqnarray}
In the following we will explore the consequences of this assumption.
The most important one is that it is possible to obtain an analytical solution of
the Kadanoff-Baym equations for the Majorana neutrino fields, which will be discussed
in the following.

\subsubsection*{Non-equilibrium Majorana Propagator}

As a first step, we derive a Kadanoff-Baym equation for the retarded and advanced propagators
\begin{eqnarray}
  \hat\pN_{R}(x,y) & = & \Theta(x^0-y^0)\hat\pN_{\rho}(x,y) \,, \nonumber\\
  \hat\pN_{A}(x,y) & = & -\Theta(y^0-x^0)\hat\pN_{\rho}(x,y) \,.
\end{eqnarray}
First, we apply the differential operator that appears on the left-hand side of the
Kadanoff-Baym Eq.\,\eqref{KBEN} to the first line,
\begin{eqnarray}
  \lefteqn{(i \slashed{\partial}_x -  M - \delta{\seN}(x)) \pN_{R}(x,y)  }\nonumber\\
  & = & i\gamma_0\delta(x^0-y^0)\pN_{\rho}(x,y) + \Theta(x^0-y^0)(i \slashed{\partial}_x -  M - \delta{\seN}(x)) \pN_{\rho}(x,y) \;.\nonumber
\end{eqnarray}
Here we have used that the spectral function at equal times is determined by the commutation
relations of the Majorana neutrino fields,
\begin{equation}
  \pN_\rho^{ij}(x,y)|_{x^0=y^0} = i \langle \left\{ N_i(x), \bar N_j(y) \right\} \rangle_{x^0=y^0} = i\gamma_0\delta^{ij}\delta(\vec{x}-\vec{y}) \;.
\end{equation}
Using the Kadanoff-Baym Eq.\,\eqref{KBEN} for the spectral function, we obtain
\begin{eqnarray}\label{SRfinitetimeinterval}
  \lefteqn{(i \slashed{\partial}_x -  M - \delta{\seN}(x)) \pN_{R}(x,y)  }\nonumber\\
  & = & -\delta^{ij}\delta(x-y) + \Theta(x^0-y^0)\int_{y^0}^{x^0} d^4z \seN_{\rho}(x,z) \pN_{\rho}(z,y) \;.
\end{eqnarray}
Next we rewrite the memory integral by using the identity
\[
  \int_{y^0}^{x^0} d^4z = \int d^4z \left[\Theta(x^0-z^0)\Theta(z^0-y^0) - \Theta(y^0-z^0)\Theta(z^0-x^0)\right] \;.
\]
where $\int d^4z=\int dz^0\int d^3z$ denotes a time-integration over the whole real axis and over spatial
coordinates. Due to the factor $\Theta(x^0-y^0)$ in Eq.\,(\ref{SRfinitetimeinterval}), only the first term contributes.
Inserting this relation yields a Kadanoff-Baym equation for the retarded propagator. The equation for the advanced propagator can be derived analogously. They can be summarized as
\begin{eqnarray}
  \lefteqn{(i \slashed{\partial}_x -  M - \delta{\seN}(x)) \pN_{R(A)}(x,y)  }\nonumber\\
  & = & -\delta^{ij}\delta(x-y) + \int d^4z \, \seN_{R(A)}(x,z) \pN_{R(A)}(z,y) \;.
\end{eqnarray}
For thermal lepton and Higgs fields, the self-energies are translation invariant and depend only on the relative coordinate $s=x-y$ (note that $\delta{\seN}(x)=\delta{\seN}(x+a)$). This implies that the equations for retarded and advanced propagators become invariant under the translation $x\to x'= x+a,\ y\to y'=y+a$. Furthermore, their initial conditions are fixed by the equal-time commutation relations, as for the spectral function. Consequently, the retarded and advanced propagators themselves are translation invariant, $\pN_{R(A)}(x,y)=\pN_{R(A)}(x',y')$. This means that they can depend only on the relative coordinate as well, i.e. $\pN_{R(A)}(x,y)=\pN_{R(A)}(x-y)$. For functions that depend on $x-y$, it is convenient to switch to momentum space. We define the Fourier transformation by
\begin{equation}
  G(x-y) = \int \frac{d^4p}{(2\pi)^4} \, e^{-ip(x-y)} G(p) \;.
\end{equation}
The Fourier transformed equation for retarded and advanced propagators reads
\kasten{
\begin{eqnarray}\label{RetAdvPropNeutrinoInThermalBath}
  \left[ \left( \slashed{p} - M_i \right)\delta^{ik} - \delta{\seN}^{ik}(p) - \seN^{ik}_{R(A)}(p) \right] \pN^{kj}_{R(A)}(p) =  -\delta^{ij} \;.
\end{eqnarray}
}
This is an algebraic equation, with Dirac as well as flavor matrix structure. The interaction with the thermal bath of lepton and Higgs is encoded in the self-energy. The sum of the retarded or advanced self-energy and the counterterm contribution yields renormalized expressions for the self-energy.

\medskip

Now we can return to the Kadanoff-Baym equations for the spectral and statistical Majorana propagators, Eq.\,\eqref{KBEN}.
The spectral function is directly related to the retarded and advanced propagators,
\begin{equation}
  \pN_\rho(x-y) = \pN_{R}(x-y) - \pN_{A}(x-y) \;.
\end{equation}
Consequently it is also translation invariant.

The Kadanoff-Baym equation  for the statistical propagator,
Eq.\,\eqref{KBEN}, can be written as
\begin{eqnarray}\label{KBENF}
  \int_0^\infty d^4z \left[ \left( \left( i \slashed{\partial}_x - M_i \right)\delta^{ik} - \delta{\seN}^{ik}(x) \right)\delta(x-z) - \seN^{ik}_{R}(x,z) \right] \pN^{kj}_{F}(z,y) \nonumber\\
  = \int_0^\infty d^4z \, \seN^{ik}_{F}(x,z)\pN^{kj}_{A}(z,y)\;.
\end{eqnarray}
In this form, the homogeneous part is written in the first line and the inhomogeneous part in the second one.
The solution can be written as the sum of a special solution of the inhomogeneous equations plus the general solution of the homogeneous equation. For the former we make the ansatz
\begin{equation}
	\pN^{ij}_{F}(x,y)_{inhom} = - \int_0^\infty d^4u \pN^{ik}_{R}(x,u) \int_0^\infty d^4z \, \seN^{kl}_{F}(u,z)\pN^{lj}_{A}(z,y) \;.
\end{equation}
The application of the operator in square brackets in Eq.\,\eqref{KBENF} will turn the retarded propagator into a delta function. Then integration over $u$ becomes trivial and yields the right-hand side of the Kadanoff-Baym equation, as required.

Next, we consider the homogeneous equation, i.e.~we set the right-hand side in Eq.\,\eqref{KBENF} to zero. The general solution will depend on the initial conditions. Since only the first derivative appears, the general solutions can be completely specified by the initial value of the propagator itself, $\pN^{ij}_{F}(x,y)|_{x^0=y^0=0}$. Here, we make the ansatz
\begin{equation}\label{HomAnsatz}
	\pN^{ij}_{F}(x,y)_{hom} = - \int d^3u \int d^3v \, \pN^{ik}_{R}(x,(0,\vec{u})) \, A^{kl}(\vec{u},\vec{v})\pN^{lj}_{A}((0,\vec{v}),y) \;,
\end{equation}
where $A^{kl}(\vec{u},\vec{v})$ is a free function.
Using again the equation of motion for the retarded propagator, one can see that this ansatz indeed solves the homogeneous equation (except for delta-functions located at the initial time surface; these determine the initial conditions but do not influence the validity of the solution). Also, it is important that the dependence on $y$ solves the hermitian conjugated Kadanoff-Baym equation, as required. The initial condition to which the ansatz corresponds can be extracted by letting $x^0,y^0\to 0$ from above. In this case we can replace the retarded and advanced functions by the spectral function, and use its equal-time limit. This yields
\begin{equation}\label{HomIC}
	\pN^{ij}_{F}(x,y)_{hom}|_{x^0=y^0=0} = i\gamma_0 A^{ij}(\vec{x},\vec{y}) i\gamma_0 \;.
\end{equation}
Thus there is a simple one-to-one correspondence between the function $A^{ij}(\vec{x},\vec{y})$ and the initial condition of the statistical propagator. The general solution is given by the sum of both contributions,
\begin{equation}
	\pN^{ij}_{F}(x,y) = \pN^{ij}_{F}(x,y)_{inhom} + \pN^{ij}_{F}(x,y)_{hom} \;.
\end{equation}
This is a formal solution, which is valid even when lepton and Higgs were not in equilibrium. However, when they are treated as a thermal bath, then the retarded and advanced functions as well as the self-energy depend only on the time-differences. Thus, in a thermal bath, the integrand in the inhomogeneous contribution involves only time-translation invariant functions. Therefore, one might expect that the inhomogeneous contribution is itself time-translation invariant. However, this is not the case because of the lower integration limits, which start at the initial time $t=0$. The deeper reason behind this is that standard Kadanoff-Baym equations rely on a hidden assumption. Namely, it is implicitly assumed that all the higher correlation functions (three-point, four-point function etc.) vanish at $t=0$, i.e. that the system is in a Gaussian state at $t=0$. Then, higher correlations have to build up, which is reflected in the finite integration range starting at $t=0$. Physically, we expect that higher correlations are present at any time. A simple way how to incorporate these is to consider also the evolution of the system at times `before' the initial time, i.e. for $t<0$. Technically, this means that all integrations start at $t=-\infty$ instead of $t=0$,
\begin{equation}
	\pN^{ij}_{F}(x,y)_{inhom} = - \int_{-\infty}^\infty d^4u \int_{-\infty}^\infty d^4v \, \left[ \pN^{ik}_{R}(x,u) \seN^{kl}_{F}(u,v)\pN^{lj}_{A}(v,y) \right]  \;.
\end{equation}
Then, the inhomogeneous contribution will no longer vanish at $x^0=y^0=0$, but will be determined by the pre-evolution of the system (i.e.~at times $x^0,y^0<0$). On the other hand, the ansatz Eq.\,\eqref{HomAnsatz} still represents a valid solution of the homogeneous equation for non-zero time arguments. It describes a Gaussian \emph{deviation} from equilibrium occurring at $x^0=y^0=0$, and can be traced back to an external source term, that is also present within the usual Gaussian framework and that originates from the contribution of the density matrix to the 2PI effective action~\cite{Garny:2009ni,Garbrecht:2011xw}.

Let us now revisit what happens when lepton and Higgs are treated as a thermal bath. Then, since retarded and advanced propagators as well as the self-energy depend only on the difference of the time arguments, and the integration range is now over the whole real axis, the inhomogeneous solution itself will be time-translation invariant, i.e.
\begin{eqnarray}
	\pN^{ij}_{F}(x,y)_{inhom} & = & - \int_{-\infty}^\infty d^4u \int_{-\infty}^\infty d^4v \, \left[ \pN^{ik}_{R}(x-u) \seN^{kl}_{F}(u-v)\pN^{lj}_{A}(v-y) \right]  \nonumber\\
	& = & \pN^{ij}_{F}(x-y)_{inhom} = \pN^{ij\,th}_{F}(x-y) \;.
\end{eqnarray}
This means that, in a thermal bath, the inhomogeneous part simply corresponds to the thermal equilibrium propagator. The deviation from equilibrium is therefore described by the homogeneous contribution Eq.\,\eqref{HomAnsatz}. In particular, from Eq.\,\eqref{HomIC} it follows that the function $A$ appearing in Eq.\,\eqref{HomAnsatz} specifies the deviation from equilibrium at the `initial' time $x^0=y^0=0$,
\begin{eqnarray}
  \Delta\pN^{ij}_{F}(x,y)|_{x^0=y^0=0} & = & (\pN^{ij}_{F}(x,y)-\pN^{ij\,th}_{F}(x-y))_{x^0=y^0=0} \nonumber\\
  & = & i\gamma_0 A^{ij}(\vec{x},\vec{y}) i\gamma_0 \;.
\end{eqnarray}
Since we are interested mainly in spatially homogeneous systems, it is convenient to switch to spatial momentum space, but to retain the dependence on time. In particular the function $A$ fulfills $A^{ij}(\vec{x},\vec{y})=A^{ij}(\vec{x}-\vec{y})$. Then the formal solution for the statistical propagator can be summarized as
\begin{eqnarray}
  \pN^{ij}_{F\,\vec{p}}(t,t') & = & - \int_{-\infty}^\infty du \int_{-\infty}^\infty dv \, \left[ \pN^{ik}_{R\,\vec{p}}(t,u) \seN^{kl}_{F\,\vec{p}}(u,v)\pN^{lj}_{A\,\vec{p}}(v,t') \right]  \nonumber\\
  && {} - \pN^{ik}_{R\,\vec{p}}(t,0) A^{ij}_{\vec{p}} \pN^{lj}_{A\,\vec{p}}(0,t') \;.
\end{eqnarray}
As discussed above, in a thermal bath the inhomogeneous part of the solution yields the thermal propagator, and the homogeneous part describes the deviation from equilibrium. The solution for the statistical propagator can thus be written as
\kasten{
\begin{eqnarray}\label{SolutionMajoranaProp}
  \pN^{ij}_{F\,\vec{p}}(t,t') & = & \pN^{ij\,th}_{F\,\vec{p}}(t-t') \nonumber\\
  && {} - \pN^{ik}_{R\,\vec{p}}(t) i\gamma_0 \Delta\pN^{kl}_{F\,\vec{p}}(0,0) i\gamma_0 \pN^{lj}_{A\,\vec{p}}(-t') \;.
\end{eqnarray}
}
Due to the retarded and advanced functions, the second line is non-zero only when $t>0$ and $t'>0$. This means that it describes the impact of a distortion that has occurred at time $t=t'=0$. In other words, it describes how the system reacts and how the distortion is propagated to positive times $t,t'>0$. We emphasize that there were no assumptions necessary that would limit the size of the `distortion' described by $\Delta\pN^{kl}_{F\,\vec{p}}(0,0)$. This means that the equation is valid also for large deviations from equilibrium, as long as the assumption of a thermal bath is justified. For example, for `zero initial abundance', the Majorana propagator at $t=t'=0$ is the vacuum one, $\pN^{kl}_{F\,\vec{p}}(0,0)=\pN^{kl\,vac}_{F\,\vec{p}}(0)$. This means one has to choose $\Delta\pN^{kl}_{F\,\vec{p}}(0,0)=\pN^{kl\,vac}_{F\,\vec{p}}(0) - \pN^{kl\,th}_{F\,\vec{p}}(0)$.

\subsubsection*{Asymmetry for thermal lepton and Higgs}

Starting point for our calculation of the lepton asymmetry is the quantum field theoretical expression Eq.\,\eqref{AsymmetryWaveContribution}. We insert the thermal lepton and Higgs propagators, and write the Majorana propagator in the form
\begin{equation}
 \pN^{ij}_{F\,\vec{p}}(t,t') = \pN^{ij\,th}_{F\,\vec{p}}(t-t') + \Delta\pN^{ij}_{F\,\vec{p}}(t,t') \;.
\end{equation}
Furthermore, we extend the lower limits of integration in Eq.\,\eqref{AsymmetryWaveContribution} to $-\infty$ in order to take higher correlations at the `initial' time $t=0$ into account, as explained above. Then the contribution to the asymmetry that involves only equilibrium propagators vanishes, because the lepton asymmetry is zero in thermal equilibrium.
Next, we insert the solution Eq.\,\eqref{SolutionMajoranaProp} of the Kadanoff-Baym equation for the statistical propagator, which can be written as
\begin{eqnarray}\label{SolutionDeltaMajoranaProp}
  \Delta\pN^{ij}_{F\,\vec{p}}(t,t') & = & - \pN^{ik}_{R\,\vec{p}}(t) i\gamma_0 \Delta\pN^{kl}_{F\,\vec{p}}(0,0) i\gamma_0 \pN^{lj}_{A\,\vec{p}}(-t') \\
  & = &  \Theta(t)\Theta(t')\pN^{ik}_{\rho\,\vec{p}}(t) i\gamma_0 \Delta\pN^{kl}_{F\,\vec{p}}(0,0) i\gamma_0 \pN^{lj}_{\rho\,\vec{p}}(-t') \;. \nonumber
\end{eqnarray}
It is non-zero only for $t>0$ and $t'>0$. This means that the lower limits of integration in Eq.\,\eqref{AsymmetryWaveContribution} are again zero. Note that it was nevertheless important to extend them in the first place, because otherwise the `equilibrium' contribution would not have vanished due to unwanted correlation build-up effects. Altogether, we arrive at the following expression for the lepton asymmetry in a thermal bath:
\kasten{
\begin{eqnarray}\label{asymmetryThermalBath}
  {n_L}_{\La\Lb}(t) & = & i\yu_{\La i }\yu_{j\Lb}^\dag \int \frac{d^3p}{(2\pi)^3} \int_0^t dt' \int_0^t dt'' \\
  && \tr \Big[ P_R \left( \Delta\pN^{ij}_{F\, \vec{p}}(t',t'') - \Delta\pNcp^{ji}_{F\, \vec{p}}(t',t'') \right) P_L \pLH_{\rho\,\vec{p}}(t''-t') \Big] \;. \nonumber
\end{eqnarray}
}

\subsubsection*{Weakly interacting thermal bath}

A case of particular interest arises when the interactions
of lepton and Higgs are strong enough to maintain thermal equilibrium,
but at the same time weak enough to justify the use of free thermal
propagators for their two-point functions. In the following we will
explore the consequences of this assumption, and we postpone the discussion
whether this assumption is justified within the SM. The
case of free thermal propagators is also interesting because it
enables a rather direct comparison with the Boltzmann result.

The free thermal propagators for the Higgs and lepton fields
in the time/momentum representation are given by
\begin{eqnarray}\label{pH}
 \pH^{eq}_{F\,\vec{k}}(t-t') & = & \frac{1+2f^{eq}_\higgs(k)}{2k}\cos\left(k(t-t')\right) \,, \nonumber\\
 \pH^{eq}_{\rho\,\vec{k}}(t-t') & = & \frac{1}{k}\sin\left(k(t-t')\right) \,,
\end{eqnarray}
and
\begin{eqnarray}\label{pL}
 \pL^{\! eq}_{F\,\vec{k}}(t-t') & = & \frac{1-2f^{eq}_\lepton(k)}{2}\left\{ \frac{-\vec{k}\cdot\vec{\gamma}}{k} \cos\left(k(t-t')\right) - i\gamma_0\sin\left(k(t-t')\right) \right\} \,, \nonumber\\
 \pL^{\! eq}_{\rho\,\vec{k}}(t-t') & = & \left\{ \frac{-\vec{k}\cdot\vec{\gamma}}{k} \sin\left(k(t-t')\right) + i\gamma_0\cos\left(k(t-t')\right) \right\} \,,
\end{eqnarray}
where $k=|\vec{k}|$, $f^{eq}_\higgs(k)=1/(e^{k/T}-1)$, and $f^{eq}_\lepton(k)=1/(e^{k/T}+1)$.
The chiral propagator is obtained by multiplying with the projector $P_L$ from left. Note that the correct chiral structure is nevertheless taken into account by the projectors $P_{L/R}$ contained in the self-energies. In addition, similar to Ref.\,\cite{Anisimov:2010dk}, we take the effect of a thermal width into account qualitatively within the
the lepton-Higgs-loop $\pLH_\rho$ appearing in Eq.~(\ref{asymmetryThermalBath})
by considering damped Breit-Wigner propagators
\begin{eqnarray}
 \pH^{eq}_{F(\rho)\,\vec{k}}(t-t') & \to & \pH^{eq}_{F(\rho)\,\vec{k}}(t-t') e^{-\Gamma_\higgs |t-t'|/2} \,, \nonumber\\
 \pL^{\! eq}_{F(\rho)\,\vec{k}}(t-t') & \to & \pL^{\! eq}_{F(\rho)\,\vec{k}}(t-t') e^{-\Gamma_\lepton |t-t'|/2}\,.
\end{eqnarray}
Evaluating the lepton-Higgs-loop $\pLH_\rho$
defined in Eq.\,(\ref{leptonHiggsLoop}) with the
thermal lepton and Higgs propagators specified above yields
\begin{eqnarray}\label{lhloop}
  \pLH^\eq_{\rho\,\vec{p}}(t-t') & = & \int \frac{d^3q}{(2\pi)^3} \Big[ \pL^{\! eq}_{\rho\,\vec{k}}(t-t') \pH^{eq}_{F\,\vec{q}}(t-t') + \pL^{\! eq}_{F\,\vec{k}}(t-t') \pH^{eq}_{\rho\,\vec{q}}(t-t') \Big]  \nonumber \\
  & = & \int \frac{d^3q}{(2\pi)^3 2q} e^{-\GammaLH |u|/2} \big\{ \pLH_{\rho}^{\!\!\!\!\!++} \, e^{i(k+q)u} + \pLH_{\rho}^{\!\!\!\!\!--} \, e^{-i(k+q)u} \\
  && \quad\quad\quad + \pLH_{\rho}^{\!\!\!\!\!+-} \, e^{i(k-q)u}+ \pLH_{\rho}^{\!\!\!\!\!-+} \, e^{-i(k-q)u}\big\} \;. \nonumber
\end{eqnarray}
 The coefficients introduced in the last line are given by
\begin{eqnarray}
 \pLH_{\rho}^{\!\!\!\!\!\pm\pm} & = & \frac{i}{2k}(k\gamma_0\pm\vec{k\gamma}) f_{\lepton\higgs}(k,q) \,, \nonumber\\
 \pLH_{\rho}^{\!\!\!\!\!\pm\mp} & = & \frac{i}{2k}(k\gamma_0\pm\vec{k\gamma}) \left( f_\higgs(q) + f_\lepton(k) \right) \,, \nonumber
\end{eqnarray}
where $\vec{k}\equiv\vec{p}-\vec{q}$, $q=|\vec{q}|$ and $k=|\vec{k}|$, $\GammaLH=\Gamma_\lepton+\Gamma_\higgs$, and
\[ f_{\lepton\higgs}(k,q) = 1 + f_\higgs(q) - f_\lepton(k) \,. \]

\section{Analytical Breit-Wigner approach }\label{sec:BW}

The equation for the lepton asymmetry, Eq.\,(\ref{asymmetryThermalBath}), together with the
expression (\ref{SolutionDeltaMajoranaProp}) for the non-equilibrium evolution of
the Majorana neutrinos in flavor space, and with the Schwinger-Dyson equations (\ref{RetAdvPropNeutrinoInThermalBath}) for the retarded and advanced neutrino propagators can be solved numerically. Before discussing the numerical solution,
it is however very useful to pursue an analytical approach. This is the content of this section.

We will proceed in three steps. First, we compute the retarded and advanced two-point functions for
the Majorana neutrinos by employing a Breit-Wigner approximation, and taking the flavor matrix structure
into account. This will allow us to identify effective pole masses and widths of the quasi-degenerate Majorana
neutrino flavors. Then, we will use these propagators to determine a solution of the Kadanoff-Baym equation for the statistical propagator, which describes the non-equilibrium evolution of the Majorana neutrino fields. Finally, this
solution will be used to compute the time-dependent lepton asymmetry produced by the
interaction of the Majorana neutrinos with the thermal lepton and Higgs background.

\subsection{Retarded and advanced propagators}

The retarded and advanced Majorana neutrino two-point functions can be obtained
by solving the algebraic Schwinger-Dyson equation (\ref{RetAdvPropNeutrinoInThermalBath})
in momentum space. Since we are interested in the time-dependence, we will also perform
a Fourier transformation.

In order to solve the Schwinger-Dyson equation (\ref{RetAdvPropNeutrinoInThermalBath}),
we insert the thermal lepton and Higgs propagators (\ref{pH},\ref{pL}) into the expression (\ref{selfEnergyCTPSeN})
for the Majorana neutrino self-energy. The self-energy can be written in the form
\begin{eqnarray}
  \seN_{R(A)}^{ij}(p) & = & -2 \left[ (h^\dag h)_{ij} P_L + (h^\dag h)_{ji} P_R \right] \pLH_{R(A)}(p) \;,
\end{eqnarray}
where we have assumed even \textit{CP} parities $\eta_i=1$ for simplicity.
The function $\pLH_{R(A)}(p)$ denotes the lepton-Higgs loop integral. It can be written as
a sum of vacuum and thermal contributions. Using dimensional regularization, the former is given by
\begin{equation}\label{LHLoop}
  \pLH_{R(A)}^{vac}(p)  =  \frac{1}{32\pi^2} \left( \frac{1}{\epsilon} -\gamma_E + \ln(4\pi) + 2 -\ln\left(\frac{|p^2|}{\mu^2}\right) \pm i\pi\Theta(p^2) \sgn(p_0)\right) \slashed{p} \;.
\end{equation}
The divergent contribution can be eliminated by introducing suitable mass and wave-function counterterms
for the Majorana neutrino fields. If the difference between the Majorana mass parameters
$M_{1,2}$ appearing in the Lagrangian is large compared to the decay width $\Gamma_{1,2}$, it is
convenient to use an on-shell renormalization scheme. However, in the present work, we are interested
in mass splittings of the order or even smaller than the width. In this case, we find it more convenient
to use $\overline{MS}$ renormalization. This means that the mass parameters $M_{1,2}$ are not necessarily
equal to the physical pole masses $M^{pole}_{1,2}$, and that the residues at the poles are not necessarily
normalized. We will derive an equation for the pole masses in the following, and also properly take
the residues into account in the computation of the lepton asymmetry. We stress that the lepton asymmetry
is independent of the renormalization scheme, and thus we can choose it to our convenience.

Accordingly, we choose the wave-function and mass counterterms as
\begin{eqnarray}
  \delta Z_{ij} & = & \frac{(h^\dag h)_{ij}}{16\pi^2}\left( \frac{1}{\epsilon} -\gamma_E + \ln(4\pi) + 2\right) \;,\\
  \delta M_{ij} & = & -\frac12 \left( M_i \delta Z_{ij} - \delta Z_{ji} M_j \right) \;.
\end{eqnarray}
In addition, it is convenient to choose the renormalization scale close to the mass scale
of the Majorana neutrinos, such that the logarithm appearing in (\ref{LHLoop}) is small.
We will set $\mu=(M_1+M_2)/2$ for definiteness. We have checked that our numerical results are
highly independent of the precise value in the range $M_1<\mu<M_2$.

In a \CP-symmetric configuration and for positive $p^0$ the thermal contribution
to the lepton-Higgs loop is given by
\begin{align}
\label{Sellphimed}
S^{th}_{\ell\phi R(A)}(p)&=\sum_{\pm}\biggl[{\int} d\Pi_k^\ell f_\ell(k_0)
\slashed{k}_\pm
{\frac{\cal P}{(p-k_\pm)^2}}-{\int} d\Pi_q^\phi f_\phi(q_0)
(p^\mu-q_\pm^\mu)\frac{\cal P}{(\slashed{p}-\slashed{q}_\pm)^2}\biggr]\nonumber\\
&\pm\frac{i}{2}\Theta(p^2)\int d\Pi_k^\ell d\Pi_q^\phi (2\pi)^4 \delta(p-q-k)\,\slashed{k}
\,\left[\,f_\phi(q_0)-f_\ell(k_0)\right]\,.
\end{align}
where, to shorten the notation, we have introduced $p_\pm=(\pm E_p,\vec p)$
and $q_\pm=(\pm E_q,\vec q)$. For negative $p^0$ the loop momentum
$k=(E_k,\vec k)$ in the second line of \eqref{Sellphimed} should be replaced with $k=(E_k,-\vec k)$. 
For the analytical treatment, we will confine ourselves to the regime for which
the Majorana neutrinos can be treated as non-relativistic when the lepton asymmetry
is produced. In this case, the thermal contribution can be approximated by
$\pLH_{R(A)}^{th}(p) \simeq -T^2\slashed{p}/(12p^2) \pm 2i\slashed{p}/(e^{|p_0|/T}-1) \Theta(p^2) \sgn(p_0)/(32\pi)$.
The first term describes a thermal mass shift, while the second one encodes the finite width. For the analytical
treatment it is convenient to change the renormalization prescription in order to take the leading
effect of the thermal mass shift into account, $\delta Z_{ij} \to \delta Z_{ij} + \delta Z_{ij}(T)$ with
$\delta Z_{ij}(T) \equiv (h^\dag h)_{ij}T^2/(6\mu^2)$. The relation between the mass- and Yukawa coupling matrices at
zero and finite temperature is 
\begin{eqnarray}
 M(T) & = & (P_LZ(T)^T+P_R Z(T)^\dag)M(T=0)(P_L Z(T) + P_R Z(T)^*)\,,\nonumber\\
 (h^\dag h)(T) & = & Z(T)^T (h^\dag h)(T=0) Z(T)^* \,,
\end{eqnarray}
where $Z_{ij}(T)\equiv V_{ik}(T)(\delta_{kj}+\delta Z_{kj}(T))$. Here $V(T)$ is a unitary
matrix in flavor space that can be adjusted such that the mass matrix at finite temperature
is diagonal and has real and positive entries, like the one at zero temperature. In the following, all quantities refer to the
finite temperature values. We stress again that the lepton asymmetry is unaffected by the field rescaling,
i.e. we would obtain identical results when setting $\delta Z_{ij}(T)$ to zero and using the vacuum parameters
instead. The rescaling is performed for computational convenience.

Altogether, when inserting the vacuum and thermal contributions to the self-energy as well as the counterterms
into the Schwinger-Dyson equation (\ref{RetAdvPropNeutrinoInThermalBath}), we obtain the following
renormalized equations for the retarded and advanced propagators:
\begin{eqnarray}\label{RetAdvPropNeutrinoInThermalBathRen}
  \left[ \left( \slashed{p} - M_i \right)\delta^{ik} + i (\gamma_{ik}P_L + \gamma_{ki} P_R) \slashed{p} \right] \pN^{kj}_{R}(p) =  -\delta^{ij} \;, \\
  \left[ \left( \slashed{p} - M_i \right)\delta^{ik} - i (\gamma^*_{ki}P_L + \gamma^*_{ik} P_R) \slashed{p} \right] \pN^{kj}_{A}(p) =  -\delta^{ij} \;.
\end{eqnarray}
The coefficients are given by 
\begin{equation}
  \gamma_{ij} = (h^\dag h)_{ij} \left[ \frac{\Theta(p^2)\sgn(p_0)}{16\pi}\left(1+\frac{2}{e^{|p_0|/T}-1}\right) + i \left(\frac{\ln\frac{|p^2|}{\mu^2}}{16\pi^2} + \frac{T^2}{6p^2} - \frac{T^2}{6\mu^2} \right)\right] \;.
\end{equation}

In the following, we will first identify the poles and
the corresponding residua of the retarded and advanced propagators,
and then Fourier transform the solution taking the
contribution from the poles into account.

\begin{figure}[t]
  \begin{center}
    \includegraphics[width=1.0\textwidth]{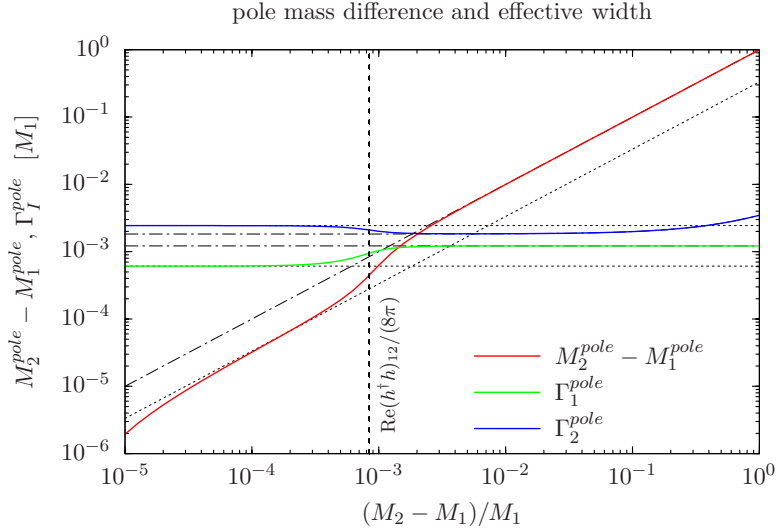}
  \end{center}
  \caption{\label{fig:spec1}
  Effective masses $M_I^{pole} \equiv \omega_{pI}|_{\vec{p}=0}$ and widths $\Gamma_I^{pole} \equiv \Gamma_{pI}|_{\vec{p}=0}$ of the sterile Majorana neutrinos extracted from the complex poles of the resummed retarded and
  advanced propagators for $(h^\dag h)_{11}=0.03$, $(h^\dag h)_{22}=0.045$, $(h^\dag h)_{12}=0.03\cdot e^{i\pi/4}$
  and $T=0.25 M_1$. The black dot-dashed lines show the approximation valid for $(M_2-M_1)/M_1\gtrsim \Re(h^\dag h)_{12}/(8\pi)$, see Eqs.~(\ref{spec_hier}), and the black dotted lines correspond to the approximate expressions (\ref{spec_deg})
  valid for $\Re(h^\dag h)_{12}/(8\pi) \gtrsim (M_2-M_1)/M_1\gg {\rm max}_{i,j}|(h^\dag h)_{ij}|^2$. }
\end{figure}

Let us first consider the retarded propagator.
In order to find a solution we make a suitable Lorentz decomposition~\cite{Buchmuller:1997yu,Plumacher:1998ex,Pilaftsis:2003gt},
\[ \pN_{R}(p) =  \pN_{LL}P_L + \pN_{RR}P_R + \pN_{LR}P_L\slashed{p} + \pN_{RL}P_R\slashed{p} \;, \]
where each coefficient is a two-by-two matrix in flavor space.
We denote the expression in square brackets on the left-hand side of the Schwinger-Dyson equation
by $\Omega$, and perform an analogous decomposition:
\begin{subequations}
\begin{align}
 \Omega_{LL}^{ik} & =   \Omega_{RR}^{ik} = - M_i\delta^{ik}  \,,  \\
 \Omega_{LR}^{ik} & =   \delta^{ik} + i \gamma_{ik} \,,  \\
 \Omega_{RL}^{ik} & =  \delta^{ik} + i \gamma_{ki} \,.
\end{align}
\end{subequations}
Then the solution of the SD equation reads:
\begin{subequations}
\begin{align}
 \pN_{LL} & =   - \left[ \Omega_{LL} - p^2 \Omega_{LR}\Omega_{RR}^{-1}\Omega_{RL} \right]^{-1}  \,, \\
 \pN_{RR} & =   - \left[ \Omega_{RR} - p^2 \Omega_{RL}\Omega_{LL}^{-1}\Omega_{LR} \right]^{-1}  \,,  \\
 \pN_{LR} & =  - \Omega_{LL}^{-1} \Omega_{LR} \pN_{RR} = \left[ \Omega_{RR}\Omega_{LR}^{-1}\Omega_{LL} - p^2 \Omega_{RL} \right]^{-1} \,,  \\
 \pN_{RL} & =   - \Omega_{RR}^{-1} \Omega_{RL} \pN_{LL} = \left[ \Omega_{LL}\Omega_{RL}^{-1}\Omega_{RR} - p^2 \Omega_{LR} \right]^{-1}  \,.  
\end{align}
\end{subequations}
The components can be computed explicitly. It is convenient to write them as
\begin{equation}
 \pN_{XY} = \frac{s_{XY}}{H} \,,
\end{equation}
where $X,Y\in L,R$, and using the common denominator $H$ given by
\begin{eqnarray}
  H & = & M_1M_2\det\left[ \Omega_{RR} - p^2 \Omega_{RL}\Omega_{LL}^{-1}\Omega_{LR} \right] \nonumber \\
  & = & M_1M_2\det\left[ \Omega_{LL} - p^2 \Omega_{LR}\Omega_{RR}^{-1}\Omega_{RL} \right] = Q^2 p^4 + M_1^2M_2^2 \nonumber\\
  &  - & p^2 \left[ M_2^2(1 + i\gamma_{11})^2 + M_1^2(1 + i\gamma_{22})^2 - M_1 M_2 (\gamma_{12}^2+\gamma_{21}^2)\right]  \,,\nonumber \\
  Q & = & \det\Omega_{LR} = \det\Omega_{RL} = (1+ i\gamma_{11})(1+ i\gamma_{22}) + \gamma_{12}\gamma_{21} \,.
\end{eqnarray}
For the enumerators, we obtain
\begin{eqnarray}\label{sXY}
 s_{LL} & = &  \left(\begin{array}{ll}
                        M_1 (M_2^2 - p^2 (1 + i\gamma_{22})^2) + M_2 p^2 \gamma_{21}^{2} \\
                        \qquad\qquad\qquad\qquad\quad i p^2 (M_1 \gamma_{12} (1 + i\gamma_{22}) + M_2  \gamma_{21} (1 + i\gamma_{11})) \\
                        i p^2 (M_1 \gamma_{12} (1 + i\gamma_{22}) + M_2  \gamma_{21}(1 + i\gamma_{11}) ) \\
                        \qquad\qquad\qquad\qquad\quad M_2 (M_1^2 - p^2 (1 + i\gamma_{11})^2) + M_1 p^2 \gamma_{12}^2
                       \end{array}\right) \,, \nonumber\\
  s_{RR} & = &  s_{LL}|_{\gamma_{12}\mapsto \gamma_{21}} \,, \nonumber\\
  s_{LR} & = & \left(\begin{array}{cc}
                        M_2^2 (1 + i \gamma_{11}) - p^2 Q (1 +i \gamma_{22}) &
                        i (M_1 M_2 \gamma_{12} + p^2 Q \gamma_{21}) \\
                        i (M_1 M_2 \gamma_{21} + p^2 Q \gamma_{12}) &
                        M_1^2 (1 + i \gamma_{22}) - p^2 Q (1 +i \gamma_{11})
                        \end{array}\right) \,, \nonumber\\
  s_{RL} & = & s_{LR}|_{\gamma_{12}\mapsto \gamma_{21}} \,.
\end{eqnarray}
The denominator $H$ can be written in the form
\begin{equation}
 H = Q^2(p^2-x_1)(p^2-x_2)\,.
\end{equation}
All the components are proportional to $1/H$ (note that $H$ is invariant under the replacement ${\gamma_{12}\mapsto \gamma_{21}}$). Therefore, the retarded propagator has
poles at $p^2=x_{1,2}$. The two complex poles are given by
\begin{equation}
 x_{1,2} = \frac{(V\pm W)^2}{4Q^2}\big|_{p^2=x_{1,2}} \,,
\end{equation}
where the quantities $V$ and $W$ have been defined as
\begin{eqnarray}
 V & = & \sqrt{ ( M_1 (1+ i\gamma_{22}) - M_2 (1+ i\gamma_{11}))^2 - M_1M_2(\gamma_{12}+\gamma_{21})^2 }\,, \nonumber\\
 W & = & \sqrt{ ( M_1 (1+ i\gamma_{22}) + M_2 (1+ i\gamma_{11}))^2 - M_1M_2(\gamma_{12}-\gamma_{21})^2 }\,. \nonumber
\end{eqnarray}
Note that these expressions are implicit equations for the poles, because the coefficients $\gamma_{ij}$
themselves depend on $p^2$. It is possible to determine the poles iteratively, starting with an
initial guess that we choose as $\gamma_{ij}(p^2=\mu^2)$ with $\mu=(M_1+M_2)/2$. 
The poles of the retarded propagator can be parameterized by effective
dispersion relations $\omega_{pI}$ and widths $\Gamma_{pI}$,
\begin{equation}
 p_0 = \pm \omega_{pI} -\frac{i}{2}\Gamma_{pI}, \quad I=1,2,
\end{equation}
which are determined by $p^2=p_0^2-\vec{p}^2=x_I$,
\begin{equation}\label{effwidthandmass}
 \Gamma_{pI} = \frac{|\Im(x_I)|}{\omega_{pI}}, \quad \omega_{pI} = \sqrt{\Re(x_I)+\Gamma_{pI}^2/4 + \vec{p}^2 } \;.
\end{equation}
The pole masses $M_I^{pole} \equiv \omega_{pI}|_{\vec{p}=0}$ and widths $\Gamma_I^{pole} \equiv \Gamma_{pI}|_{\vec{p}=0}$
obtained for vanishing spatial momentum are shown in Fig.~\ref{fig:spec1} for a representative choice of
parameters. We remind the reader that the masses $M_{1,2}$ are the renormalized mass parameters in the
modified $\overline{MS}$ scheme where the mass shift due to the finite temperature is already absorbed
into $M_{1,2}$. The relation to the vacuum mass parameters can be easily inferred from the rescaling
prescription discussed above. 

It is instructive to derive approximate analytical expressions in two limiting cases of
interest. Since we are mainly interested in mass splittings of the order of the width,
$(M_2-M_1)/M_1 \sim \mathcal{O}((h^\dag h)_{ij})$, and small Yukawa couplings
$|(h^\dag h)_{ij}|\ll 1$, we will assume in all cases that
$(M_2-M_1)/M_1 \gg {\rm max}_{i,j}|(h^\dag h)_{ij}|^2$. We also assume that the different entries
are not fine-tuned, such that e.g. the difference $(h^\dag h)_{22}-(h^\dag h)_{11}$ is
of the same order of magnitude as the individual terms. Provided these assumptions are
satisfied, we find that simple approximate expressions can be obtained depending on
the relative size of $(M_2-M_1)/M_1$ and the real part of the flavor
off-diagonal combination $\Re(h^\dag h)_{12}/(8\pi)$ of Yukawa couplings.
Let us first consider the regime $(M_2-M_1)/M_1 \gtrsim \Re(h^\dag h)_{12}/(8\pi)$. Here we find that
\begin{eqnarray}\label{spec_hier}
  M_i^{pole} & \simeq & M_i \;, \nonumber \\
  \Gamma_i^{pole} & \simeq & \Gamma_i \equiv \frac{(h^\dag h)_{ii}}{8\pi}M_i\left(1+\frac{2}{e^{M_i/T}-1}\right) \;.
\end{eqnarray}
Thus, in this case, the pole masses are approximately equal to the renormalized mass parameters,
and the widths are given by the same expressions as in the hierarchical case.

If, on the other hand, $(M_2-M_1)/M_1 \lesssim \Re(h^\dag h)_{12}/(8\pi)$, we find
\begin{eqnarray}\label{spec_deg}
  M_i^{pole}  & \simeq & \frac{M_1+M_2}{2}\pm\frac{(M_2-M_1) ((h^\dag h)_{22}-(h^\dag h)_{11})}{2\sqrt{((h^\dag h)_{22}-(h^\dag h)_{11})^2+4(\Re(h^\dag h)_{12})^2}} \;, \nonumber \\
  \Gamma_i^{pole} & \simeq & \frac{M_i}{16\pi}\left(1+\frac{2}{e^{M_i/T}-1}\right)\Big( (h^\dag h)_{11} + (h^\dag h)_{22} \nonumber \\
&& {} \pm \sqrt{((h^\dag h)_{22}-(h^\dag h)_{11})^2+4(\Re(h^\dag h)_{12})^2}\Big) \;.
\end{eqnarray}
Note that the off-diagonal coupling parameter is bounded from above,
$|(h^\dag h)_{12}|^2 \leq (h^\dag h)_{11} (h^\dag h)_{22}$, such that the effective width
cannot become negative. As can be seen in Fig.~\ref{fig:spec1}, the transition
between the two regimes is relatively fast, i.e. the approximate expressions can be used for
most of the considered parameter space. In addition, both cases agree if $\Re(h^\dag h)_{12}/((h^\dag h)_{22}-(h^\dag h)_{11})\ll 1$ is small. We find that Eq.~(\ref{spec_hier}) yields a reasonable approximation within both regimes
if the ratio is   smaller than $\sim 1/3$.

\medskip

It turns out that the replacement $\gamma_{ij} \to \gamma_{ij}(p^2=\mu^2)$
yields a rather accurate approximation for small Yukawa couplings and mass splittings of the order
or larger than the width. We will use this simplification in the following in order to obtain an analytical solution
for the retarded propagator. In addition, we will assume that $1+2/(e^{|p_0|/T}-1) \approx 1$ for simplicity. The resulting solution can be brought into the form
\begin{equation}
 \pN_{R}(p) = \frac{Z_{1R}}{p^2-x_1} + \frac{Z_{2R}}{p^2-x_2}  \,,
\end{equation}
where the residua $Z_{IR}$ are matrices in flavor space, and can be decomposed analogously to the propagator,
\[ Z_{IR} = Z_{ILL} P_L + Z_{IRR} P_R + Z_{ILR} P_L\slashed{p} + Z_{IRL} P_R\slashed{p} \quad\mbox{for}\ I=1,2 \;. \]
The components $Z_{IXY}$ ($X,Y=L,R$) of the residues can be computed explicitly using
\begin{equation}
 Z_{1XY} = \frac{1}{Q^2}\frac{s_{XY}(p^2=x_1)}{x_1-x_2},\quad
 Z_{2XY} = \frac{1}{Q^2}\frac{s_{XY}(p^2=x_2)}{x_2-x_1} \;.
\end{equation}

The solution for the advanced propagator can be formally obtained by replacing $i\gamma \to (i\gamma)^\dag$,
which implies $\Omega_{XY}\to \Omega_{XY}^\dag$. Using that $\Omega_{LL}$ and $\Omega_{RR}$ are hermitian, we find that the advanced propagator is given by
\begin{equation}
 \pN_A(p) =  \pN_{RR}^\dag P_L + \pN_{LL}^\dag P_R + \pN_{LR}^\dag P_L\slashed{p} + \pN_{RL}^\dag P_R\slashed{p} \;,
\end{equation}
where all expressions are evaluated at momentum $p$.
For the poles, hermitian conjugation is equivalent to $x_I\mapsto x_I^*$. Therefore, we obtain
\begin{equation}
 \pN_{A}(p) = \frac{Z_{1A}}{p^2-x_1^*} + \frac{Z_{2A}}{p^2-x_2^*} \,,
\end{equation}
where $ Z_{IA} =  Z_{IRR}^\dag P_L + Z_{ILL}^\dag P_R + Z_{ILR}^\dag P_L\slashed{p} + Z_{IRL}^\dag P_R\slashed{p} $
for $I=1,2$. 

We are interested in the retarded and advanced propagators in time rather than
frequency representation. Both representations are related by the Fourier transformation
\begin{equation}
 \pN_{R(A)\,\vec{p}}(t) = \int \frac{dp_0}{2\pi} \, e^{-ip_0t} \, \pN_{R(A)}(p)  \,.
\end{equation}

The Fourier integration can be performed using the Cauchy theorem, and closing the contour in the upper imaginary plane for $t<0$ and in the lower one for $t>0$. Note that, due to the non-analytic behaviour of $\gamma_{ij}$ at the threshold $p^2=0$,
it is necessary to use a contour which is obtained by dividing a semi-circle into three parts such that the
lines corresponding to $\Re(p_0)=\pm|\vec{p}|$ are left out. Within the analytical Breit-Wigner approach pursued
here, we will take only the contributions from the poles into account, and neglect the contributions
that arise from the parts of the contours parallel to the imaginary axis. This will be justified later on by
comparison with numerical solutions. Since the poles of the retarded propagator have a negative imaginary part, the integral vanishes in the case $t<0$. This is precisely what is expected for a retarded propagator. Analogously, the advanced propagator vanishes when $t>0$. We find for the result
\begin{eqnarray}\label{retadvBW}
 \pN_{R(A)\,\vec{p}}(t) & = & \pm\Theta(\pm t) \sum_{I=1,2} e^{\mp\Gamma_{pI}t/2} \Big[ e^{-i\omega_{pI}t}\pN_{IR(A)}^-  + e^{+i\omega_{pI}t}\pN_{IR(A)}^+ \Big]  \;,
\end{eqnarray}
where, the positive and negative frequency coefficients for retarded and advanced function, respectively, are given by
\begin{eqnarray}\label{retadvcoeff}
 \pN_{IR}^\pm & = & \pm\frac{i}{2} \frac{Z_{IR}|_{p^0=\mp\omega_{pI}- i\Gamma_{pI}/2}}{\omega_{pI}\pm i\Gamma_{pI}/2} \,, \nonumber\\
 \pN_{IA}^\pm & = & \pm\frac{i}{2} \frac{Z_{IA}|_{p^0=\mp\omega_{pI}+ i\Gamma_{pI}/2}}{\omega_{pI}\mp i\Gamma_{pI}/2} \,. 
\end{eqnarray}
In the limit of small Yukawa couplings and mass splitting larger than the width, the retarded and advanced propagators approach the well-known result
\begin{equation}
 \pN_{R(A)\,\vec{p}}^{ij}(t) \to \pm\Theta(\pm t) e^{\mp\Gamma_{pi}t/2} 
 \Big[ \frac{M_i-\vec{p\gamma}}{\omega_{pi}} \sin(\omega_{pi}t) +i\gamma_0\cos(\omega_{pi}t)\Big]\delta^{ij}\;.
\end{equation}
The flavour off-diagonal contributions, which are suppressed by the Yukawa coupling, are important for
the creation of a lepton asymmetry. The leading contributions obtained by expanding in the
 Yukawa coupling are given in the Appendix.

\subsection{Lepton asymmetry}

The statistical propagator for the Majorana neutrinos, which describes the
non-equilibrium time evolution due to the interaction with the thermal lepton and Higgs
background, can be directly obtained by inserting the retarded and advanced propagators 
into Eq.~(\ref{SolutionDeltaMajoranaProp}). Next, we obtain the lepton asymmetry
generated by the deviation of the Majorana neutrinos from equilibrium from
Eq.~(\ref{asymmetryThermalBath}). It is convenient to rewrite it in the form
\begin{eqnarray}
  n_L(t) & = & \int \frac{d^3p}{(2\pi)^3}  \tr \Big[ \Delta\pN^{kl}_{F\,\vec{p}}(0,0)K^{lk}_{\vec{p}}(t) - \Delta\pNcp^{kl}_{F\,\vec{p}}(0,0)\bar K^{lk}_{\vec{p}}(t)\Big] \;. \nonumber
\end{eqnarray}
The lepton asymmetry depends on the initial conditions for the Majorana neutrino,
and on the time-dependent coefficients $K$ which are given by
\begin{eqnarray}
  K_{\vec{p}}^{lk}(t) & \equiv & i(\yu^\dag\yu)_{ji} \int_0^t dt' \int_0^t dt''\gamma_0 \pN^{lj}_{A\,\vec{p}}(-t'') P_L \pLH_{\rho\,\vec{p}}(t''-t') P_R \pN^{ik}_{R\,\vec{p}}(t')\gamma_0 \,, \nonumber\\
  \bar K^{lk}_{\vec{p}}(t) & \equiv & i(\yu^\dag\yu)_{ij} \int_0^t dt' \int_0^t dt''\gamma_0 \pNcp^{lj}_{A\,\vec{p}}(-t'') P_L \pLH_{\rho\,\vec{p}}(t''-t') P_R \pNcp^{ik}_{R\,\vec{p}}(t')\gamma_0 \;. \nonumber
\end{eqnarray}
In order to compute the asymmetry, we insert Eq.~(\ref{lhloop}) for the lepton-Higgs loop $\pLH_{\rho\,\vec{p}}(t''-t')$.
Using the result for the retarded and advanced propagators, we can now decompose the double-time integral,
\begin{eqnarray}
  K_{\vec{p}}^{lk}(t) & = &  \int \frac{d^3q}{(2\pi)^3 2q}\sum_{I,J=1,2}\sum_{\epsilon_k=\pm 1} \gamma^{lk}_{JI}(\{\epsilon_k\}) \, L_{IJ}(t,\{\epsilon_k\})\,, \nonumber\\
  \bar K_{\vec{p}}^{lk}(t) & = &  \int \frac{d^3q}{(2\pi)^3 2q}\sum_{I,J=1,2}\sum_{\epsilon_k=\pm 1} \bar \gamma^{lk}_{JI}(\{\epsilon_k\}) \, L_{IJ}(t,\{\epsilon_k\})\,,
\end{eqnarray}
where all the $\epsilon_k$ are summed over $+$ and $-$ modes. The coefficients are given by
\begin{eqnarray}
  \gamma^{lk}_{JI}(\{\epsilon_k\}) & = & -i(\yu^\dag\yu)_{ji} \gamma_0 \pN^{lj\epsilon_1}_{AJ} P_L \pLH_{\rho\,\vec{p}}^{\epsilon_2\epsilon_3} P_R \pN^{ik\epsilon_4}_{RI}\gamma_0 \,, \nonumber\\
  \bar \gamma^{lk}_{JI}(\{\epsilon_k\}) & = & -i(\yu^\dag\yu)_{ij} \gamma_0 \pNcp^{lj\epsilon_1}_{AJ} P_L \pLH_{\rho\,\vec{p}}^{\epsilon_2\epsilon_3} P_R \pNcp^{ik\epsilon_4}_{RI}\gamma_0 \,,
\end{eqnarray}
and the time integrals are explicitly given by
\begin{equation}
 L_{IJ}(t,\{\epsilon_k\}) = \int_0^t dt' \int_0^t dt'' \, e^{iAt''} \, e^{-iBt'} \, e^{-\GammaLH |t''-t'|/2}
\end{equation}
where
\begin{eqnarray}
 A & = & \epsilon_2k + \epsilon_3q - \epsilon_1\omega_{pJ} + i\Gamma_{pJ}/2 \,, \nonumber\\
 B & = & \epsilon_2k + \epsilon_3q - \epsilon_4\omega_{pI} - i\Gamma_{pI}/2 \,.
\end{eqnarray}
The double time integral can be easily performed now. Since the interesting dynamics occurs at time scales of $t \sim 1/\Gamma_{pI}$, and the width of lepton and Higgs is much larger than the one of the Majorana neutrinos, it is reasonable to consider the limit where
\begin{equation}
 \GammaLH \cdot t \gg 1 \;.
\end{equation}
In this limit,
\begin{eqnarray}
 L_{IJ}(t,\{\epsilon_k\}) & \simeq & \frac{i\GammaLH}{2(A-B)}\left(\frac{1}{A^2+\GammaLH^2/4}+\frac{1}{B^2+\GammaLH^2/4}\right)\left[1-e^{i(A-B)t}\right] \nonumber\\
 && + \frac{AB-\GammaLH^2/4}{(A^2+\GammaLH^2/4)(B^2+\GammaLH^2/4)}\left[1+e^{i(A-B)t}\right] \;.
\end{eqnarray}
Lets identify which terms yield the leading contribution to the lepton asymmetry. Formally, we have to look for contributions which can be of order $1/\Gamma_1$, $1/\Gamma_2$ or $1/(M_1-M_2)$. Only the term proportional to $1/(A-B)$ can be of that order, provided that $\epsilon_1=\epsilon_4$. Then for $I=J$ we get a contribution of order $1/\Gamma_I$, and for $I\not= J$ of order $1/(\omega_{p1}-\omega_{p2} + i(\Gamma_{p1}+\Gamma_{p2})/2 )$. The latter will be important in the maximal resonant case. Thus, from now on we only take the $1/(A-B)$ term into account.

Furthermore, the terms involving $A^2$ or $B^2$ are unsuppressed only if the corresponding real parts of A or B can become zero within the phase-space integration. This is only the case for $\epsilon_2=\epsilon_3=\epsilon_1$ or $\epsilon_2=\epsilon_3=\epsilon_4$, respectively. Altogether, this means the leading contribution arises when all $\epsilon$'s are identical,
\begin{equation}
 L_{IJ}(t,\{\epsilon_k\}) \quad \to \quad L_{IJ}^\epsilon(t) \equiv L_{IJ}(t,\{\epsilon_k\})|_{\epsilon_1=\epsilon_2=\epsilon_3=\epsilon_4\equiv\epsilon} \;.
\end{equation}
Then, we find the following result:
\begin{align}
& L_{IJ}^\epsilon(t)  =   \frac{i\epsilon}{\omega_{pI}-\omega_{pJ}+i\epsilon(\Gamma_{pI}+\Gamma_{pJ})/2} \left[ 1 - e^{i\epsilon(\omega_{pI}-\omega_{pJ})t-(\Gamma_{pI}+\Gamma_{pJ})t/2} \right] \nonumber\\
 &  \times \left( \frac{\GammaLH/2}{(\omega_{pJ}-k-q+i\epsilon\Gamma_{pJ}/2)^2+\GammaLH^2/4} + \frac{\GammaLH/2}{(\omega_{pI}-k-q-i\epsilon\Gamma_{pI}/2)^2+\GammaLH^2/4}\right) \nonumber
\end{align}
The entries on the diagonal can be written in a simpler form. They are real-valued and equal for both
signs of $\epsilon$, $L_{II}^+(t)=L_{II}^-(t)\equiv L_{II}(t)$, with
\begin{eqnarray}
 L_{II}(t) & = & \frac{1 - e^{-\Gamma_{pI}t} }{\Gamma_{pI}}\,  \Re \left( \frac{\GammaLH}{(\omega_{pI}-k-q+i\Gamma_{pI}/2)^2+\GammaLH^2/4} \right) \,.
\end{eqnarray}
Note that the time-dependence precisely coincides with the time-evolution of the
lepton asymmetry that would be expected from
a Boltzmann approach.

The off-diagonal entries, i.e. for $I\not= J$, lead to an oscillatory behaviour and will be important in the maximal resonant regime. They fulfill the relations
\[ L_{12}^\pm(t) = L_{21}^\pm(t)^* = L_{12}^\mp(t)^* \]

Note that all entries except $L_{11}$ become suppressed in the hierarchical limit $M_2\sim\omega_{p2}\gg M_1$. On the other hand, in the extreme degenerate limit $M_2\to M_1$, $\Gamma_2\to \Gamma_1$, all the entries $L_{IJ}$ become equal.

Putting the results together, we obtain the following expression for the lepton asymmetry:
\begin{eqnarray}
  n_L(t) & = & \int \frac{d^3p}{(2\pi)^3}  \int \frac{d^3q}{(2\pi)^3 2q}\sum_{I,J=1,2}\sum_{\epsilon=\pm 1} F_{JI}^\epsilon L_{IJ}^\epsilon(t)  \;,
\end{eqnarray}
where
\begin{eqnarray}\label{FIJ}
  F_{JI}^\epsilon & = &  \tr \Big[ \Delta\pN^{kl}_{F\,\vec{p}}(0,0)\gamma^{lk\epsilon}_{JI} - \Delta\pNcp^{kl}_{F\,\vec{p}}(0,0)\bar \gamma^{lk\epsilon}_{JI} \Big] \;,
\end{eqnarray}
and
\begin{eqnarray}
  \gamma^{lk\epsilon}_{JI} & = & \gamma^{lk}_{JI}(\{\epsilon_k=\epsilon\}) = -i(\yu^\dag\yu)_{ji} \gamma_0 \pN^{lj\epsilon}_{AJ} P_L \pLH_{\rho\,\vec{p}}^{\epsilon\epsilon} P_R \pN^{ik\epsilon}_{RI}\gamma_0 \,, \nonumber\\
  \bar \gamma^{lk\epsilon}_{JI}  & = & \bar\gamma^{lk}_{JI}(\{\epsilon_k=\epsilon\}) = -i(\yu^\dag\yu)_{ij} \gamma_0 \pNcp^{lj\epsilon}_{AJ} P_L \pLH_{\rho\,\vec{p}}^{\epsilon\epsilon} P_R \pNcp^{ik\epsilon}_{RI}\gamma_0 \,.
\end{eqnarray}
Using the properties of $L_{IJ}^\epsilon(t)$, one can write the lepton asymmetry as
a sum of Boltzmann-like and oscillatory contributions,
\kasten{
\begin{eqnarray}\label{LeptonAsymmetryBreitWigner}
  n_L(t) & = & \int \frac{d^3p}{(2\pi)^3}  \int \frac{d^3q}{(2\pi)^3 2q}\Big[ F_{11} L_{11}(t) + F_{22} L_{22}(t) \nonumber\\
  && {} + F_{12}^s \Re(L_{21}^+(t) ) + F_{12}^a \Im(L_{21}^+(t) ) \Big] \;,
\end{eqnarray}
}
where the coefficients $F$ contain all the information about the interaction, the \textit{CP} violation
and the initial state,
\begin{eqnarray}
  F_{II} & \equiv & F_{II}^+ + F_{II}^- \,, \nonumber \\
  F_{12}^s & \equiv & F_{12}^+ + F_{12}^- + F_{21}^+ + F_{21}^- \,,  \\
  F_{12}^a & \equiv & i \, \left( F_{12}^+ - F_{12}^- - F_{21}^+ + F_{21}^- \right)\,. \nonumber
\end{eqnarray}
and the functions $L(t)$ characterize the time-evolution,
\kasten{
\begin{eqnarray}\label{LIJ}
 L_{II}(t) & = & \frac{1 - e^{-\Gamma_{pI}t} }{\Gamma_{pI}}\,  \Re \left( \frac{\GammaLH}{(\omega_{pI}-k-q+i\Gamma_{pI}/2)^2+\GammaLH^2/4} \right) \,, \nonumber\\
 L_{21}^+(t)  & = & \frac{1 - e^{-i(\omega_{p1}-\omega_{p2})t}e^{-\Gamma_{p}t}}{\Gamma_{p}+i(\omega_{p1}-\omega_{p2})} \Bigg( \frac{\GammaLH/2}{(\omega_{p1}-k-q+i\Gamma_{p1}/2)^2+\GammaLH^2/4} \nonumber\\
 &&  + \frac{\GammaLH/2}{(\omega_{p2}-k-q-i\Gamma_{p2}/2)^2+\GammaLH^2/4} \Bigg) \;.
\end{eqnarray}
where $\Gamma_p\equiv(\Gamma_{p1}+\Gamma_{p2})/2$.
}

One non-trivial element that remains to be specified is the initial condition for the statistical neutrino propagator $\Delta\pN_{F\,\vec{p}}(0,0)$. In principle, we have a great freedom to impose initial conditions, as far as the validity of the approach is concerned. Clearly, it is desirable to motivate the choice by some physical considerations. Concretely, we will assume that the neutrinos are in `vacuum' at $t=t'=0$, i.e.
\begin{equation}
 \pN_{F\,\vec{p}}(t,t')|_{t=t'=0} = \pN^{vac}_{F\,\vec{p}}(t-t')|_{t=t'=0} \;.
\end{equation}
The Majorana neutrinos will then be produced by the Yukawa interactions with the thermal bath of Higgs boson and leptons. This resembles a typical initial condition often used for leptogenesis calculations.

By definition, the above choice implies that the deviation from equilibrium at the initial time is given by
\begin{equation}
 \Delta\pN_{F\,\vec{p}}(t,t')|_{t=t'=0} = \pN^{vac}_{F\,\vec{p}}(t-t')|_{t=t'=0} - \pN^{th}_{F\,\vec{p}}(t-t')|_{t=t'=0}\;.
\end{equation}

\subsection{Expansion in the Yukawa couplings}\label{sec:expansion}

By combining the results for the lepton asymmetry (\ref{LeptonAsymmetryBreitWigner})
and the coefficients (\ref{retadvcoeff}) of the positive and negative frequency
modes of the retarded and advanced propagators, it is possible to determine the
resonantly generated asymmetry for a given set of Yukawa couplings $(h^\dag h)_{ij}$ and
mass parameters $M_{1,2}$. The approximations done so far allow to choose a mass
difference of the order of the width, $(M_2-M_1)/M_1 \sim \mathcal{O}((h^\dag h)_{ij})/(8\pi) \ll 1$,
which is the regime in which we expect a maximal enhancement. It is possible to simplify
the analytical expressions considerably within the parameter region $\Re(h^\dag h)_{12} \ll |(h^\dag h)_{22}-(h^\dag h)_{11}|$
and $(M_2-M_1)/M_1 \gtrsim \Re(h^\dag h)_{12}/(8\pi)$. We will therefore first discuss the analytical
solution in this regime, and then compare it to the full analytical solution in Breit-Wigner
approximation as well as a full numerical solution.

In the parameter region described above, the pole masses are approximately equal to
the renormalized mass parameters $M_{1,2}$ and the widths are equal to the usual
expression for the decay rate $\Gamma_{1,2}$ given in Eq.~(\ref{spec_hier}).
The dispersion relation and damping rate of the retarded and advanced propagators are
given by $\omega_{pI}\simeq \sqrt{M_I^2+\vec{p}^2}$ and $\Gamma_{pI}\simeq \Gamma_IM_I/\omega_{pI}$.
Then, it is possible to obtain explicit analytical expressions for the coefficients
$F_{IJ}$ defined in Eq.~(\ref{FIJ}) by systematically expanding the solutions (\ref{sXY})
for retarded and advanced propagators for small Yukawa couplings $(h^\dag h)_{ij} \ll 1$.
In particular, at leading order in the Yukawa couplings, the vacuum initial condition is given by
\begin{equation}\label{ICLO}
  \Delta\pN_{F\,\vec{p}}^{ij}(t,t')|_{t=t'=0} \simeq \delta^{ij} f_{FD}(\omega_{pi})\frac{M_i-\vec{p\gamma}}{\omega_{pi}}\;.
\end{equation}
The result for the coefficients $F_{IJ}$ is shown in the Appendix, including also several intermediate steps of
the calculation. The result obtained for the lepton asymmetry will be discussed
in the following section.

\section{Result for the lepton asymmetry}\label{sec:result}

In this section, we will first compare our analytical result for the lepton
asymmetry obtained within the Kadanoff-Baym approach in Breit-Wigner approximation with
the one obtained from a Boltzmann treatment. Then, we will compare various 
analytical, semi-analytical and numerical results obtained within the Kadanoff-Baym framework.

\subsection{Comparison of Kadanoff-Baym and Boltzmann}

Let us start with discussing the explicit analytical result in Breit-Wigner
approximation, and compare it to the well-known Boltzmann approach. 
More specifically, we will first present the result that
is obtained under the assumptions described in section \ref{sec:expansion}.
By inserting the coefficients $F_{IJ}$ obtained at leading order in the
Yukawa couplings into the expression (\ref{LeptonAsymmetryBreitWigner})
for the lepton asymmetry, we obtain the following result for its time-evolution:
\kasten{
\begin{eqnarray}
  n_L(t) & = & \frac{\Im[(\yu^\dag\yu)_{12}^2]}{8\pi} \int \frac{d^3p}{(2\pi)^3} \int \frac{d^3q}{(2\pi)^3 2q} \int \frac{d^3k}{(2\pi)^3 2k} (2\pi)^3\delta(\vec{p}-\vec{k}-\vec{q})\nonumber\\
  && \times  (1+f_{\higgs}(q)-f_{\lepton}(k)) \nonumber\\
  && \times \Bigg[ \sum_{I=1,2} \frac{4 k\cdot p_I }{\omega_{pI}}  f_{FD}(\omega_{pI})\, \Re\left(\frac{M_1M_2\, L_{II}(t)}{M_2^2-M_1^2- iM_1\Gamma_1+ iM_2\Gamma_2}\right) \nonumber\\
  && -  \frac{4 k\cdot p_2}{\omega_p}\, f_{FD}(\omega_{p1})\Re\left(\frac{M_1M_2\,L_{21}^+(t)}{M_2^2-M_1^2 - iM_1\Gamma_1 + iM_2\Gamma_2}\right) \nonumber\\
  && -  \frac{4 k\cdot p_1}{\omega_p}\, f_{FD}(\omega_{p2})\Re\left(\frac{M_1M_2\,L_{21}^+(t)^*}{M_2^2-M_1^2 - iM_1\Gamma_1 + iM_2\Gamma_2}\right)\Bigg] \;. \nonumber
\end{eqnarray}
}
where $q=|\vec{q}|$ and $k=|\vec{k}|$ denote the Higgs and lepton energies, $\omega_{pI}=\sqrt{M_I^2+\vec{p}^2}$ are the energies of the Majorana neutrinos ($I=1,2$), and $k\cdot p_I \equiv k\omega_{pI}-\vec{kp}$. We have also defined $\omega_p\equiv 2\omega_{p1} \omega_{p2}/(\omega_{p1}+\omega_{p2})$. The time-dependence is described by the functions $L(t)$. Neglecting Yukawa-suppressed contributions in the expressions (\ref{LIJ}) yields
\kasten{
\begin{eqnarray}
 L_{II}(t) & = & \frac{1 - e^{-\Gamma_{pI}t} }{\Gamma_{pI}}\,   \frac{\GammaLH}{(\omega_{pI}-k-q)^2+\GammaLH^2/4}  \,, \nonumber\\
 L_{21}^+(t)  & = & \frac{1 - e^{-i(\omega_{p1}-\omega_{p2})t}e^{-\Gamma_{p}t}}{\Gamma_{p}+i(\omega_{p1}-\omega_{p2})} \Big( \frac{\GammaLH/2}{(\omega_{p1}-k-q)^2+\GammaLH^2/4} \nonumber\\
 &&  + \frac{\GammaLH/2}{(\omega_{p2}-k-q)^2+\GammaLH^2/4} \Big) \;,\nonumber
\end{eqnarray}
}
where $\GammaLH=\Gamma_\lepton+\Gamma_\higgs$ is the sum of the thermal width
of lepton and Higgs, and $\Gamma_{pI}=M_I\Gamma_I/\omega_{pI}$ is related to the decay widths
of the Majorana neutrinos. Furthermore, we have defined $\Gamma_p\equiv(\Gamma_{p1}+\Gamma_{p2})/2$.

The flavor-diagonal contributions $L_{11}(t)$ and $L_{22}(t)$ feature a time-dependence that is also expected within
the Boltzmann approach for the contribution from the neutrino species $N_1$ and $N_2$, respectively. 
In addition, there
exists a flavor off-diagonal contribution $L_{21}^+(t)$, which has an oscillating behaviour with frequency given by
the energy difference $\omega_{p1}-\omega_{p2}$ of the two
Majorana neutrino species. The latter can be interpreted as the contribution from coherent transitions
between the neutrino mass eigenstates $N_1$ and $N_2$ due to the off-diagonal Yukawa coupling $\yu^\dag\yu_{12}$. 

\medskip

It is instructive to compare the result for the lepton asymmetry with the classical Boltzmann
approach. Within the latter, the lepton asymmetry is determined by the Boltzmann equation
\begin{eqnarray}
  \frac{d n_L^{\it Boltzmann}}{dt} & = & \sum_{I=1,2}\int \frac{d^3p}{(2\pi)^3\omega_{pI}} \int \frac{d^3q}{(2\pi)^3 2q} \int \frac{d^3k}{(2\pi)^3 2k} (2\pi)^4\delta(p_I-k-q)\nonumber\\
  && \times\, \, \epsilon^{CP}_I|{\cal M}_{N_I\to\lepton\higgs}|^2\Big[ f_{N_I}(1+f_{\higgs})(1-f_{\lepton})
- (1-f_{N_I})f_{\higgs}f_{\lepton}\Big] \;, \nonumber
\end{eqnarray}
where 
\[ \epsilon^{CP}_I = \frac{1}{(\yu^\dag\yu)_{II}} \times \frac{\Im[(\yu^\dag\yu)_{IJ}^2]}{8\pi}\, \frac{M_IM_J(M_J^2-M_I^2)}{(M_J^2-M_I^2)^2+(M_I\Gamma_J - M_J\Gamma_J)^2} \;, \]
is the `wave' contribution to the \textit{CP}-violating parameter \cite{Anisimov:2005hr} (see also \cite{Pilaftsis:2003gt}), and $|{\cal M}_{N_I\to\lepton\higgs}|^2=4k\cdot p_I(\yu^\dag\yu)_{II}$ is related to the tree-level matrix element for the decay $N_I\to\lepton\higgs$. For the setup considered here, the time-evolution of the classical distribution function is simply given by $f_{N_I}(t)=f_{FD}(\omega_{pI})(1-e^{-\Gamma_{pI}t})$.

In the following, we will compare the Kadanoff-Baym with the Boltzmann result in the hierarchical and in the degenerate
limit of the Majorana neutrino spectrum.

\subsubsection*{Hierarchical limit}

In the hierarchical limit $M_2\gg M_1$, the contributions proportional to the Fermi-Dirac distribution
evaluated at the energy of $N_2$, $f_{FD}(\omega_{p2})$,
are exponentially suppressed for the relevant temperature range $T\sim M_1$. In addition, the coherent
contribution $L_{21}^+(t) \propto 1/(\Gamma_{p}+i(\omega_{p1}-\omega_{p2})) \sim i/\omega_{p2}$ is
strongly suppressed compared to the flavor-diagonal contribution $L_{11}(t)\propto 1/\Gamma_{p1}$.
The lepton asymmetry obtained in the Kadanoff-Baym approach is thus given by
\begin{eqnarray}
  n_L(t) & = & \frac{\Im[(\yu^\dag\yu)_{12}^2]}{8\pi}\frac{M_1}{M_2} \int \frac{d^3p}{(2\pi)^3\omega_{p1}} \int \frac{d^3q}{(2\pi)^3 2q} \int \frac{d^3k}{(2\pi)^3 2k} \nonumber\\
  && \times\, {} (2\pi)^3\delta(\vec{p}-\vec{k}-\vec{q})\, \times\, \frac{\GammaLH}{(\omega_{p1}-k-q)^2+\GammaLH^2/4}\nonumber\\
  && \times\, {} 4 k\cdot p_1\,(1+f_{\higgs}(q)-f_{\lepton}(k))\, f_{FD}(\omega_{p1})  \frac{1 - e^{-\Gamma_{p1}t} }{\Gamma_{p1}} \;. \nonumber
\end{eqnarray}
For comparison, within the classical Boltzmann approach one finds that 
\begin{eqnarray}
  n_L^{\it Boltzmann}(t) & = & \frac{\Im[(\yu^\dag\yu)_{12}^2]}{8\pi}\frac{M_1}{M_2} \int \frac{d^3p}{(2\pi)^3\omega_{p1}} \int \frac{d^3q}{(2\pi)^3 2q} \int \frac{d^3k}{(2\pi)^3 2k} \nonumber\\
  && \times\, {} (2\pi)^3\delta(\vec{p}-\vec{k}-\vec{q})\, \times\, 2\pi\delta(\omega_{p1}-k-q)\nonumber\\
  && \times\, {} 4 k\cdot p_1\,(1+f_{\higgs}(q)-f_{\lepton}(k))\, f_{FD}(\omega_{p1})  \frac{1 - e^{-\Gamma_{p1}t} }{\Gamma_{p1}} \;. \nonumber
\end{eqnarray}
Thus, the thermal width of lepton and Higgs $\GammaLH=\Gamma_\lepton+\Gamma_\higgs$ leads to a replacement of the on-shell
delta function in the Boltzmann equations by a Breit-Wigner curve, in accordance with \cite{Anisimov:2010dk},
 \[ 2\pi\delta(\omega_{p}-k-q) \to \frac{\GammaLH }{(\omega_{p}-k-q)^2+\GammaLH^2/4}\;. \]

\subsubsection*{Degenerate limit}

In the limit where the masses are quasi-degenerate, $|M_1-M_2| \ll M_{1,2}$,
we approximate $\omega_{p1}\approx\omega_{p2}\approx\omega_p$ in all terms except
for those containing the difference of the energies. Then we obtain
\kasten{
\begin{eqnarray}\label{nLKBdeg}
  n_L(t) & = & \frac{\Im[(\yu^\dag\yu)_{12}^2]}{8\pi}\, \frac{M_1M_2(M_2^2-M_1^2)}{(M_2^2-M_1^2)^2+(M_1\Gamma_1 - M_2\Gamma_2)^2}\nonumber\\
  && \times  \int \frac{d^3p}{(2\pi)^3\omega_{p1}} \int \frac{d^3q}{(2\pi)^3 2q} \int \frac{d^3k}{(2\pi)^3 2k} \nonumber\\
  && \times\, {} (2\pi)^3\delta(\vec{p}-\vec{k}-\vec{q})\, \times\, \frac{\GammaLH}{(\omega_{p}-k-q)^2+\GammaLH^2/4}\nonumber\\
  && \times 4 k\cdot p \, (1+f_{\higgs}(q)-f_{\lepton}(k)) f_{FD}(\omega_{p})\\
  && \times \Bigg[ \sum_{I=1,2} \frac{1 - e^{-\Gamma_{pI}t} }{\Gamma_{pI}}  - 4\, \Re\left(\frac{1 - e^{-i(\omega_{p1}-\omega_{p2})t}e^{-(\Gamma_{p1}+\Gamma_{p2})t/2}}{\Gamma_{p1}+\Gamma_{p2}+2i(\omega_{p1}-\omega_{p2})}\right)\Bigg] \;. \nonumber
\end{eqnarray}
}
For comparison, the Boltzmann result in the degenerate limit reads
\begin{eqnarray}
  n_L^{\it Boltzmann}(t) & = & \frac{\Im[(\yu^\dag\yu)_{12}^2]}{8\pi}\, \frac{M_1M_2(M_2^2-M_1^2)}{(M_2^2-M_1^2)^2+(M_1\Gamma_1 - M_2\Gamma_2)^2}\nonumber\\
  && \times  \int \frac{d^3p}{(2\pi)^3\omega_{p1}} \int \frac{d^3q}{(2\pi)^3 2q} \int \frac{d^3k}{(2\pi)^3 2k} \nonumber\\
  && \times\, {} (2\pi)^3\delta(\vec{p}-\vec{k}-\vec{q})\, \times\, 2\pi\delta(\omega_{p}-k-q)\\
  && \times 4 k\cdot p \, (1+f_{\higgs}(q)-f_{\lepton}(k)) f_{FD}(\omega_{p}) \Bigg[ \sum_{I=1,2} \frac{1 - e^{-\Gamma_{pI}t} }{\Gamma_{pI}}  \Bigg] \;. \nonumber
\end{eqnarray}
The first line of both results agrees, and corresponds to the resonant enhancement described by the `usual' \textit{CP}-violating
parameter $\epsilon^{CP}_i$. Note in particular that in the denominator the `regulator' $M_1\Gamma_1 - M_2\Gamma_2$ appears,
in accordance with \cite{Anisimov:2005hr}. Furthermore, as in the hierarchical case, the on-shell delta function for the energies in the Boltzmann result is replaced by a Breit-Wigner curve in the Kadanoff-Baym result.

Finally, the Kadanoff-Baym result features an additional contribution inside the square brackets in the last line as compared to the
Boltzmann result. It results from the flavor off-diagonal contribution $L_{21}^+(t)$. As discussed above, it can be attributed to
the contribution to the lepton asymmetry from coherent flavor transitions between the Majorana neutrino species. It becomes
comparable in size with the Boltzmann-like contributions when the energy difference $\omega_{p1}-\omega_{p2}$ is of the order
of the average of the decay rates, $(\Gamma_{p1}+\Gamma_{p2})/2$, of the Majorana neutrinos. This case occurs when the
mass difference $\Delta M=M_2-M_1$ of the Majorana neutrino mass eigenstates is of the order of the widths $\Gamma_I$.
Such a situation has been frequently considered, because it leads to the maximal possible resonant enhancement of the \textit{CP} asymmetry.

\begin{figure}[t]
  \begin{center}
    \includegraphics[width=0.9\textwidth]{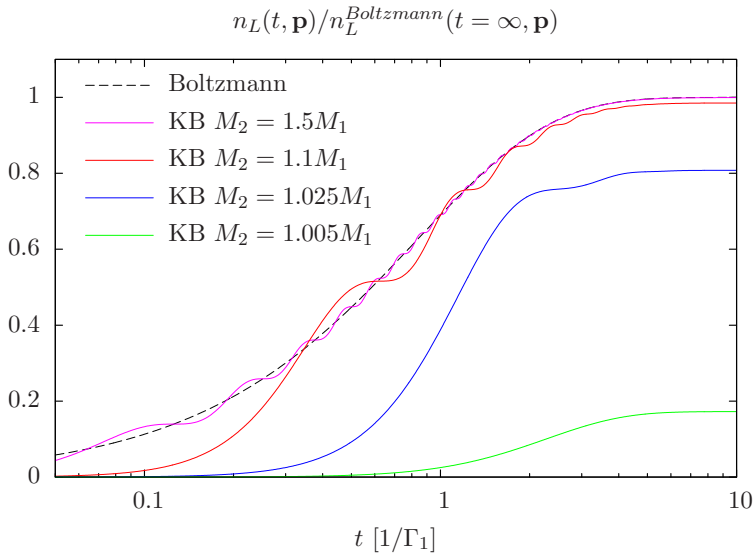}
  \end{center}
  \caption{\label{fig:fig1}
  Time-evolution of the lepton asymmetry $n_L(t,\vec{p})$. Shown is the Boltzmann result (dashed) and the result obtained
  from the Kadanoff-Baym equations, for various values of the mass-squared splitting $M_2^2-M_1^2$ (solid lines). All curves are normalized such that the Boltzmann result approaches unity. For the widths we have chosen $\Gamma_1=0.01M_1$, $\Gamma_2=0.015M_1$, and $\vec{p}=0$, for illustration.}
\end{figure}

The time-evolution of the lepton asymmetry $n_L(t,\vec{p})$, defined via $n_L(t)\equiv \int d^3p/(2\pi)^3\,n_L(t,\vec{p})$, is
shown in figure \ref{fig:fig1} for various choices of the mass-squared splitting $M_2^2-M_1^2$. When the splitting is very large
compared to the product $M_I\Gamma_I$ ($I=1,2$), the Kadanoff-Baym solution oscillates around the Boltzmann result, and the
value at large times is very similar for both cases. When the splitting is of comparable size as the product of the widths
and the masses, there are significant differences. Finally, when $M_2^2-M_1^2 \lesssim |M_2\Gamma_2-M_1\Gamma_1|$, the
lepton asymmetry obtained in the Kadanoff-Baym analysis is suppressed compared to the one predicted by the Boltzmann approach.
The reason is that the coherent contribution has the tendency to cancel part of the Boltzmann-like contributions to the
lepton asymmetry. Note that to obtain figure \ref{fig:fig1} we have neglected the effect of the thermal width of lepton
and Higgs. Their inclusion would multiply the Kadanoff-Baym results for $n_L(t,\vec{p})$ by an additional overall factor,
that is identical to the ratio of lepton asymmetries obtained from Kadanoff-Baym and Boltzmann analyses in the hierarchical limit.

In order to quantify the suppression of the lepton asymmetry that results from the additional coherent contributions,
it is useful to investigate the parametric dependence on the mass-splitting, and to identify under which conditions the
maximal possible resonant enhancement occurs.
Within the Boltzmann approach, the resonant enhancement of the lepton asymmetry is maximized provided that the function
\begin{equation}
  R^{\it Boltzmann} \equiv \frac{M_1M_2(M_2^2-M_1^2)}{(M_2^2-M_1^2)^2+(M_1\Gamma_1 - M_2\Gamma_2)^2} \;,
\end{equation}
is maximally large. We may assume $M_2>M_1$ without loss of generality. The maximum of the function $R^{\it Boltzmann}$ occurs for a mass splitting of $M_2^2-M_1^2 = |M_1\Gamma_1-M_2\Gamma_2|$,
i.e. when the mass difference is of the order of the difference of the decay widths of the neutrinos.
The maximum is given by
\begin{equation}
  R^{\it Boltzmann}_{\it max} = \frac{M_1M_2}{2|M_1\Gamma_1-M_2\Gamma_2|} \;.
\end{equation}
Since the decay widths of the neutrinos are suppressed by the Yukawa couplings, this
ratio can be much larger than one and can compensate the one-loop suppression of the \textit{CP}-violating rates,
so that it is possible to lower the scale of leptogenesis considerably below $10^9$GeV \cite{Pilaftsis:2003gt}.
In addition, it has been noticed that apparently a further enhancement occurs when not only the masses of the
neutrinos are quasi-degenerate, but also their decay widths $\Gamma_1$ and $\Gamma_2$ are of similar size,
so that $R_{\it max} \propto |M_1\Gamma_1-M_2\Gamma_2|^{-1}$ is further enhanced. Phenomenologically, this situation
could occur when the degeneracy of the masses and of the couplings originates from an underlying
flavor symmetry (see e.g. \cite{Blanchet:2009kk}).

\begin{figure}[t]
  \begin{center}
    \includegraphics[width=0.9\textwidth]{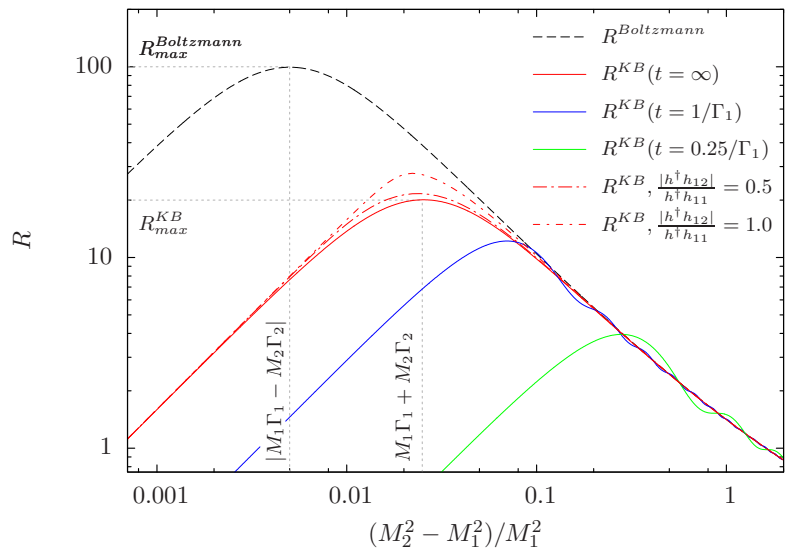}
  \end{center}
  \caption{\label{fig:fig2}
  Function $R$ characterizing the amount of resonant enhancement. Shown is the dependence on the mass-squared splitting $M_2^2-M_1^2$ for the Boltzmann result (dashed) and for the result obtained
  based on Kadanoff-Baym equations, for three fixed times $t=0.25/\Gamma_1,1/\Gamma_1,\infty$ (solid lines). The widths are chosen as in figure \ref{fig:fig1}. The dotted lines indicate the maximal resonant enhancement as predicted by the Boltzmann and Kadanoff-Baym results, respectively.  The dot-dashed lines shows the numerical result obtained for large
off-diagonal Yukawa couplings $|h^\dag h|_{12}/(h^\dag h)_{11}=0.5$ and $1$, respectively. Here $(h^\dag h)_{ii}$
are chosen as for the analytical result, and $\arg(h^\dag h)_{12}=\pi/8$. The numerical and analytical results agree for $|h^\dag h|_{12}\ll |(h^\dag h)_{22}-(h^\dag h)_{11}|$, as expected.}
\end{figure}

The Kadanoff-Baym result for the lepton asymmetry is proportional to the function $R^{\it Boltzmann}$ as well. However, the additional coherent contributions to the lepton asymmetry have the tendency to cancel the Boltzmann-like contribution. Therefore,
in order to determine the maximal possible enhancement, it is useful to define a modified, time-dependent generalization by
\begin{equation}
  R^{\it KB}(t) \equiv R^{\it Boltzmann} \times \frac{\sum\limits_{I=1,2} \frac{1 - e^{-\Gamma_{pI}t} }{\Gamma_{pI}}  - 4\, \Re\left(\frac{1 - e^{-i(\omega_{p1}-\omega_{p2})t}e^{-(\Gamma_{p1}+\Gamma_{p2})t/2}}{\Gamma_{p1}+\Gamma_{p2}+2i(\omega_{p1}-\omega_{p2})}\right)}{ \sum\limits_{I=1,2} \frac{1 - e^{-\Gamma_{pI}t} }{\Gamma_{pI}} }\;.
\end{equation}

The dependence of $R^{\it KB}(t)$ and $R^{\it Boltzmann}$ on the mass-squared splitting $M_2^2-M_1^2$ is shown in figure \ref{fig:fig2}, for a choice of parameters for which $\Gamma_2/\Gamma_1=1.5$. We observe that the resonant enhancement obtained
in the Kadanoff-Baym analysis is smaller compared to the Boltzmann approach, in accordance with the above discussion.
In order to identify the maximal possible enhancement, it is instructive to consider the value of $R^{\it KB}(t)$ for times $t\gtrsim 1/\Gamma_{pI}$,
\begin{eqnarray}
  R^{\it KB}(t)|_{t\gtrsim 1/\Gamma_{pI}} & = & R^{\it Boltzmann} \times \frac{(\omega_{p1}-\omega_{p2})^2+(\Gamma_{p1}-\Gamma_{p2})^2/4}{(\omega_{p1}-\omega_{p2})^2+(\Gamma_{p1}+\Gamma_{p2})^2/4} \nonumber\\
 & \simeq & R^{\it Boltzmann} \times \frac{(M_2^2-M_1^2)^2+(M_1\Gamma_1-M_2\Gamma_2)^2}{(M_2^2-M_1^2)^2+(M_1\Gamma_1+M_2\Gamma_2)^2} \;. \nonumber\\
\end{eqnarray}
In the last line we have used that for quasi-degenerate masses $\Gamma_{pI}=M_I\Gamma_I/\omega_{pI}\simeq M_I\Gamma_I/\omega_p$ and $\omega_{p1}-\omega_{p2} \simeq (M_1^2-M_2^2)/(2\omega_p)$.
Thus, compared to the Boltzmann treatment, the coherent contributions
lead to an additional factor that effectively changes the regulator in the resonant term from the difference to the
sum of the decay rates. This means the maximal enhancement of the lepton asymmetry occurs in fact for $M_2^2-M_1^2 = M_1\Gamma_1+M_2\Gamma_2$, and is given by
\begin{equation}
  R^{\it KB}_{\it max} = \frac{M_1M_2}{2(M_1\Gamma_1+M_2\Gamma_2)} \;.
\end{equation}
Consequently, the coherent contributions have the effect to cut off the resonant enhancement that occurs in the
doubly degenerate limit $M_1\to M_2$ and $\Gamma_1\to\Gamma_2$ within the Boltzmann approach. 

As is apparent from the expressions derived above, the maximal resonant enhancement within the Kadanoff-Baym
analysis is always smaller compared to the Boltzmann approach. The predictions of both approaches are comparable
if either $\Gamma_2/\Gamma_1\ll 1$ or $\Gamma_2/\Gamma_1\gg 1$, while the Kadanoff-Baym result is significantly smaller
than the Boltzmann result when both decay widths are of comparable size.

\subsection{Comparison of analytical and numerical results}

The explicit analytical solutions obtained in the Kadanoff-Baym approach that have been
discussed in this section have been obtained by performing a Breit-Wigner approximation
and an expansion for small Yukawa couplings that is valid
within the parameter region $\Re(h^\dag h)_{12} \ll |(h^\dag h)_{22}-(h^\dag h)_{11}|$
and $(M_2-M_1)/M_1 \gtrsim \Re(h^\dag h)_{12}/(8\pi)$.

We will now compare the analytical results with a semi-analytical and then with a numerical approach.
The semi-analytical result is also based on the Breit-Wigner approximation. 
However, the coefficients $F_{IJ}$ entering in the expression (\ref{LeptonAsymmetryBreitWigner})
for the lepton asymmetry are computed using the full solutions for the effective masses and
widths $\omega_{pI}$ and $\Gamma_{pI}$ given in Eq.~(\ref{effwidthandmass}) as well as the
full expressions (\ref{sXY}) for the retarded and advanced propagators. Consequently, this
semi-analytical approach is valid also when $\Re(h^\dag h)_{12} \sim (h^\dag h)_{ii}$
and for mass splittings $(M_2-M_1)/M_1$ larger or smaller than $\Re(h^\dag h)_{12}/(8\pi)$.
We have checked that the results obtained in both approaches agree
provided that $|h^\dag h|_{12}\ll |(h^\dag h)_{22}-(h^\dag h)_{11}|$.

The semi-analytical result is shown as dot-dashed lines in Fig.~\ref{fig:fig2} for two choices
of $|h^\dag h|_{12}$. The resonant enhancement becomes slightly stronger when $|h^\dag h|_{12}$
is increased. However, the mass splitting for which the maximal enhancement occurs remains
approximately the same. In addition, a good agreement is found also for extremely small mass
splittings. Formally the validity of the analytical approach requires that
$(M_2-M_1)/M_1 \gtrsim \Re(h^\dag h)_{12}/(8\pi) \sim 0.005(0.01)$ for
$|h^\dag h|_{12}/(h^\dag h)_{11}=0.5(1)$. Nevertheless, the comparison with the
semi-analytical approach indicates that the asymptotic value of the resonant
enhancement is predicted correctly also for extremely small mass splittings. It is worth
noting that the semi-analytical result can be reproduced rather accurately 
by replacing the first line of Eq.~(\ref{nLKBdeg}) by
\begin{eqnarray}
 \frac{\Im[(\yu^\dag\yu)_{12}^2]}{8\pi}\, \frac{M_1M_2(M_2^2-M_1^2)}{((M_2^{pole})^2-(M_1^{pole})^2)^2+(M^{pole}_1\Gamma^{pole}_1 - M^{pole}_2\Gamma^{pole}_2)^2} \;.
\end{eqnarray}
and using the full expressions for $\omega_{pI}$ and $\Gamma_{pI}$.
Here the denominator contains the pole masses and widths discussed in section~\ref{sec:BW},
and the enumerator is proportional to the basis-invariant quantity
\[ \Im[\tr[h^\dag h M^\dag M M^\dag h^Th^* M]] = \Im[(\yu^\dag\yu)_{12}^2]M_1M_2(M_2^2-M_1^2) \,,\]
related to \textit{CP}-violation \cite{Pilaftsis:1997jf}. Since the pole masses and widths agree with
the mass and width parameters in the domain of validity of the analytical result, see
Eq.~(\ref{spec_hier}), this replacement is consistent with the results discussed previously.
The corresponding generalization of the enhancement factor leads to 
\[ R^{\it KB}(t)|_{t\gtrsim 1/\Gamma_{pI}} \simeq \frac{M_1M_2(M_2^2-M_1^2)}{((M_2^{pole})^2-(M_1^{pole})^2)^2+(M^{pole}_1\Gamma^{pole}_1 + M^{pole}_2\Gamma^{pole}_2)^2} \]
This expression reproduces the numerical results shown in Fig.~\ref{fig:fig2} with very good
accuracy. It coincides with the expression obtained from the analytical solution for mass
splittings that are large compared to $\Re(h^\dag h)_{12}/(8\pi)$, and also in the
opposite limit, because then $\Gamma^{pole}_1 + \Gamma^{pole}_2 \simeq \Gamma_1 + \Gamma_2$,
as can be seen from Eq.~(\ref{spec_deg}). This behaviour is apparent also in Fig.~\ref{fig:fig2}.

\medskip

In order to cross-check the Breit-Wigner approximation, we have also solved the Kadanoff-Baym
equation (\ref{KBENspec}) for the spectral function of the Majorana neutrinos directly in time representation,
for the spatial momentum mode $\vec p = 0$.
The contribution from this mode to the lepton asymmetry can then be obtained via Eqs.(\ref{asymmetryThermalBath}, \ref{SolutionDeltaMajoranaProp}). We have used the one-loop expression (\ref{selfEnergyCTPSeN}) for the self-energy of the Majorana neutrinos, and inserted the thermal lepton and Higgs propagators (\ref{pH},\ref{pL}). For the numerical
calculation it is convenient to use an exponential momentum cutoff $e^{-|\vec{q}|/\Lambda}$ for the computation
of the lepton-Higgs loop integral, and to solve the equation for the bare propagators. The renormalization is
then performed by Fourier transforming the regulated loop integral to frequency space, and applying a
field renormalization such that the resulting expression matches with the
renormalized Schwinger-Dyson Eq.~(\ref{RetAdvPropNeutrinoInThermalBathRen}). For the cutoff we have
used values in the range $\Lambda/M_1 \sim 3.3-5$. We have checked that our renormalized prescription removes the dependence on the cutoff at a satisfactory level. Furthermore, we have used a time step $a_t=0.025/M_1$ and computed the
solution for temperatures $T\to 0$ and $T/M_1=0.25$ in the limit $\Gamma_\ell,\Gamma_\phi\to 0$. For the renormalized mass and coupling parameters we have chosen the same values that are shown in Fig.~\ref{fig:fig1}. We find that the numerical
solution for the spectral function can be well described by a sum of exponentially damped oscillating functions
with damping rate and frequency in accordance with the Breit-Wigner solution (\ref{retadvBW}). In addition,
the resulting lepton asymmetry agrees well with the analytical result for $|h^\dag h|_{12}\ll |(h^\dag h)_{22}-(h^\dag h)_{11}|$
and with the semi-analytical result for $\Re(h^\dag h)_{12} \sim (h^\dag h)_{ii}$.

\section{Conclusions}\label{sec:conclusion}

In this work we have studied the resonant enhancement of the lepton asymmetry that is generated 
by the lepton-number violating interaction of lepton and Higgs fields with a pair of quasi-degenerate Majorana neutrino species within a first-principle approach. The quantum-mechanical treatment of this system allows to study the resonant enhancement
in the case when the mass difference of the heavy neutrino species is comparable to their width. This regime is
of particular interest, because it corresponds to the parameter region for which the resonant enhancement is maximal,
and thus determines the amount by which the efficiency of leptogenesis can be increased compared to the
case of a hierarchical spectrum of the Majorana neutrinos. By treating the lepton and Higgs fields as a thermal
bath and neglecting the backreaction, it was possible to obtain a formal analytical solution for the
Kadanoff-Baym equations that describe the non-equilibrium evolution of the coupled system of Majorana neutrinos.
We have used this solution in order to determine the generation of a lepton asymmetry, for which we have
obtained approximate analytical as well as numerical results. 
Finally, we have compared the resonant enhancement
that is obtained from the quantum-mechanical Kadanoff-Baym approach to the classical Boltzmann approach.

We find that for mass splittings of the Majorana neutrinos that are large compared to their widths
the lepton asymmetries obtained within the Kadanoff-Baym and the Boltzmann approach have the
same time-dependence, and differ by an overall factor that is related to the finite width of
lepton and Higgs fields. When the mass splitting of the Majorana neutrinos is only slightly larger than their
widths, the Kadanoff-Baym
result is given by a sum of Boltzmann-like contributions and oscillating contributions which
can be attributed to coherent transitions between the two Majorana neutrino species. If the mass
splitting is of the same order as the width, a cancellation of the Boltzmann-like and
coherent contributions occurs in the Kadanoff-Baym result and the lepton asymmetry is
suppressed compared to the Boltzmann result. We have also
compared the analytical approximation of the Kadanoff-Baym result to numerical results, and found good
agreement within the domain of validity. In addition, we find that the qualitative
behaviour is robust.

Our analytical result confirms the form of the regulator given in Eq.~(\ref{EpsilonAnisimov}). However,
the cancellation that arises due to the coherent flavor transitions reduces the
maximal possible enhancement.  The analytical treatment indicates that the resonant enhancement
can be estimated approximately by replacing the `regulator' according to $M_1\Gamma_1 - M_2\Gamma_2 \to M_1\Gamma_1 + M_2\Gamma_2$. The cancellation is particularly important when
the widths of the two neutrino species are of comparable size. We therefore expect that the coherent
flavor transitions are crucial
within phenomenological scenarios for which the masses and the decay rates are quasi-degenerate. 
However, since the interplay of
Boltzmann-like and coherent contributions is a dynamical process, it is necessary to investigate these scenarios in detail.
Depending on the considered scenario, it may be appropriate to consider 
initial condition for the Majorana neutrinos that are different from the vacuum initial conditions considered here,
which will also have an impact on the asymmetry. In addition, we note that it would be interesting to
include the back-reaction and to implement the gauge interactions of lepton and Higgs fields in a systematic way.
Nevertheless, we expect that the partial cancellation of the final asymmetry due to coherence effects is a generical result.

\section*{Acknowledgements}

MG is grateful to Wilfried Buchm\"uller, Marco Drewes, Clemens Kiessig,
Sebastian Mendizabal, Markus Michael M\"uller and Christoph Weniger for
valuable discussions and comments. The work of MG was partially supported by the DFG
cluster of excellence ``Origin and Structure of the Universe.''
AK is supported by DFG under Grant KA-3274/1-1 ``Systematic analysis of baryogenesis in non-equilibrium quantum field theory''. 

\begin{appendix}

\section{Expansion in the Yukawa couplings}

In this Appendix approximate analytical expressions
for the coefficients $F_{IJ}$ defined in Eq.~(\ref{FIJ}) are derived, under the assumptions discussed in section~\ref{sec:expansion}. These coefficients determine the
resonant enhancement of the lepton asymmetry $n_L(t)$that is generated
by Majorana neutrinos interacting with a thermal bath, according to Eq.~(\ref{LeptonAsymmetryBreitWigner}).

Within the parameter region described in section~\ref{sec:expansion}, the complex poles of the
retarded and advanced Majorana neutrino two-point functions obtained at leading order
in the Yukawa couplings are given by Eq.~(\ref{spec_hier}),
\begin{eqnarray}
  x_I & = & M_I^2 - iM_I\Gamma_I, \quad I=1,2\,.
\end{eqnarray}
For the residues of the poles we find in the same approximation,
\begin{eqnarray}
  Z_{1R} & = & - \left(\begin{array}{cc}
                        M_1+\slashed{p}  & 0 \\ 0 & 0
                        \end{array}\right) + \frac{1}{x_1-x_2}\Bigg\{ i(M_1^2-M_2^2)\frac{\Gamma_1}{2M_1}\left(\begin{array}{cc}
                        2M_1+\slashed{p}  & 0 \\ 0 & 0
                        \end{array}\right) \nonumber\\
          && + i(M_1+M_2)(M_2\Gamma_1-M_1\Gamma_2)\frac{\tan(2\theta)}{4M_2}\left(\begin{array}{cc}
                        0 & M_1+\slashed{p} \\ M_1+\slashed{p} & 0
                        \end{array}\right) \nonumber\\
          && + (\Im\gamma_{12})M_1(M_1-M_2)\left(\begin{array}{cc}
                        0 & \gamma_5( M_1-\slashed{p}) \\ \gamma_5(M_1+\slashed{p}) & 0
                        \end{array}\right)\Bigg\} \,,\\
  Z_{2R} & = & - \left(\begin{array}{cc}
                        0 &  0 \\ 0 & M_2+\slashed{p} 
                        \end{array}\right) + \frac{1}{x_1-x_2}\Bigg\{ i(M_1^2-M_2^2)\frac{\Gamma_2}{2M_2}\left(\begin{array}{cc}
                        0 &  0 \\ 0 &2M_2+\slashed{p} 
                        \end{array}\right) \nonumber\\
          && - i(M_1+M_2)(M_2\Gamma_1-M_1\Gamma_2)\frac{\tan(2\theta)}{4M_1}\left(\begin{array}{cc}
                        0 & M_2+\slashed{p} \\ M_2+\slashed{p} & 0
                        \end{array}\right) \nonumber\\
          && - (\Im\gamma_{12})M_2(M_1-M_2)\left(\begin{array}{cc}
                        0 & \gamma_5( M_2+\slashed{p}) \\ \gamma_5(M_2-\slashed{p}) & 0
                        \end{array}\right)\Bigg\} \,.
\end{eqnarray}
The positive and negative frequency coefficients for Fourier transformed
retarded and advanced function can then be obtained using Eq.~(\ref{retadvcoeff}).
At zeroth order in the Yukawa couplings, the coefficients reduce to
\begin{eqnarray}
 \pN_{1R}^- & \to  & \pN_{1A}^- \to \frac{i}{2} \frac{M_1+ \omega_{p1}\gamma_0 - \vec{p\gamma} }{\omega_{p1}}\left({1\atop 0} {0\atop 0} \right) \,, \nonumber \\
 \pN_{1R}^+ & \to & \pN_{1A}^+ \to -\frac{i}{2} \frac{M_1 -\omega_{p1}\gamma_0 - \vec{p\gamma}}{\omega_{p1}}\left({1\atop 0} {0\atop 0} \right) \,, \nonumber 
\end{eqnarray}
and the result for $I=2$ is analogous. The leading contributions to the off-diagonal elements
arise at first order in $h^\dag h$ and are given by
\begin{eqnarray}
  \pN_{1R}^{12-} & = & \frac{M_1}{2\omega_{p1}} \frac{  (M_1+\omega_{p1}\gamma_0-\vec{p\gamma} ) [ \Delta\Gamma - \frac{i}{16\pi} \Im( h^\dag h_{12})(M_1-M_2)\gamma_5 ] }{M_1^2-M_2^2-iM_1\Gamma_1+iM_2\Gamma_2} \,, \nonumber \\
  \pN_{1R}^{21-} & = & \frac{M_1}{2\omega_{p1}} \frac{  [ \Delta\Gamma - \frac{i}{16\pi} \Im( h^\dag h_{12})(M_1-M_2)\gamma_5 ](M_1+\omega_{p1}\gamma_0-\vec{p\gamma} )  }{M_1^2-M_2^2-iM_1\Gamma_1+iM_2\Gamma_2} \,, \nonumber \\
  \pN_{1R}^{12+} & = & \frac{M_1}{2\omega_{p1}} \frac{  (M_1-\omega_{p1}\gamma_0-\vec{p\gamma} ) [ \Delta\Gamma - \frac{i}{16\pi} \Im( h^\dag h_{12})(M_1-M_2)\gamma_5 ] }{M_1^2-M_2^2+iM_1\Gamma_1-iM_2\Gamma_2} \,, \nonumber \\
  \pN_{1R}^{21+} & = & \frac{M_1}{2\omega_{p1}} \frac{  [ \Delta\Gamma - \frac{i}{16\pi} \Im( h^\dag h_{12})(M_1-M_2)\gamma_5 ](M_1-\omega_{p1}\gamma_0-\vec{p\gamma} )  }{M_1^2-M_2^2+iM_1\Gamma_1-iM_2\Gamma_2} \,, \nonumber 
\end{eqnarray}
\begin{eqnarray}
  \pN_{1A}^{12-} & = & -\frac{M_1}{2\omega_{p1}} \frac{  (M_1+\omega_{p1}\gamma_0-\vec{p\gamma} ) [ \Delta\Gamma - \frac{i}{16\pi} \Im( h^\dag h_{12})(M_1-M_2)\gamma_5 ] }{M_1^2-M_2^2+iM_1\Gamma_1-iM_2\Gamma_2} \,, \nonumber \\
  \pN_{1A}^{21-} & = & -\frac{M_1}{2\omega_{p1}} \frac{  [ \Delta\Gamma - \frac{i}{16\pi} \Im( h^\dag h_{12})(M_1-M_2)\gamma_5 ](M_1+\omega_{p1}\gamma_0-\vec{p\gamma} )  }{M_1^2-M_2^2+iM_1\Gamma_1-iM_2\Gamma_2} \,, \nonumber \\
  \pN_{1A}^{12+} & = & -\frac{M_1}{2\omega_{p1}} \frac{  (M_1-\omega_{p1}\gamma_0-\vec{p\gamma} ) [ \Delta\Gamma - \frac{i}{16\pi} \Im( h^\dag h_{12})(M_1-M_2)\gamma_5 ] }{M_1^2-M_2^2-iM_1\Gamma_1+iM_2\Gamma_2} \,, \nonumber \\
  \pN_{1A}^{21+} & = & -\frac{M_1}{2\omega_{p1}} \frac{  [ \Delta\Gamma - \frac{i}{16\pi} \Im( h^\dag h_{12})(M_1-M_2)\gamma_5 ](M_1-\omega_{p1}\gamma_0-\vec{p\gamma} )  }{M_1^2-M_2^2-iM_1\Gamma_1+iM_2\Gamma_2} \,, \nonumber 
\end{eqnarray}
\begin{eqnarray}
  \pN_{2R}^{12-} & = & -\frac{M_2}{2\omega_{p2}} \frac{  [ \Delta\Gamma - \frac{i}{16\pi} \Im( h^\dag h_{12})(M_1-M_2)\gamma_5 ](M_2+\omega_{p2}\gamma_0-\vec{p\gamma} )  }{M_1^2-M_2^2-iM_1\Gamma_1+iM_2\Gamma_2} \,, \nonumber \\
  \pN_{2R}^{21-} & = & -\frac{M_2}{2\omega_{p2}} \frac{  (M_2+\omega_{p2}\gamma_0-\vec{p\gamma} ) [ \Delta\Gamma - \frac{i}{16\pi} \Im( h^\dag h_{12})(M_1-M_2)\gamma_5 ] }{M_1^2-M_2^2-iM_1\Gamma_1+iM_2\Gamma_2} \,, \nonumber \\
  \pN_{2R}^{12+} & = & -\frac{M_2}{2\omega_{p2}} \frac{  [ \Delta\Gamma - \frac{i}{16\pi} \Im( h^\dag h_{12})(M_1-M_2)\gamma_5 ](M_2-\omega_{p2}\gamma_0-\vec{p\gamma} )  }{M_1^2-M_2^2+iM_1\Gamma_1-iM_2\Gamma_2} \,, \nonumber \\
  \pN_{2R}^{21+} & = & -\frac{M_2}{2\omega_{p2}} \frac{  (M_2-\omega_{p2}\gamma_0-\vec{p\gamma} ) [ \Delta\Gamma - \frac{i}{16\pi} \Im( h^\dag h_{12})(M_1-M_2)\gamma_5 ] }{M_1^2-M_2^2+iM_1\Gamma_1-iM_2\Gamma_2} \,, \nonumber 
\end{eqnarray}
\begin{eqnarray}
  \pN_{2A}^{12-} & = & \frac{M_2}{2\omega_{p2}} \frac{  [ \Delta\Gamma - \frac{i}{16\pi} \Im( h^\dag h_{12})(M_1-M_2)\gamma_5 ](M_2+\omega_{p2}\gamma_0-\vec{p\gamma} )  }{M_1^2-M_2^2+iM_1\Gamma_1-iM_2\Gamma_2} \,, \nonumber \\
  \pN_{2A}^{21-} & = & \frac{M_2}{2\omega_{p2}} \frac{  (M_2+\omega_{p2}\gamma_0-\vec{p\gamma} ) [ \Delta\Gamma - \frac{i}{16\pi} \Im( h^\dag h_{12})(M_1-M_2)\gamma_5 ] }{M_1^2-M_2^2+iM_1\Gamma_1-iM_2\Gamma_2} \,, \nonumber \\
  \pN_{2A}^{12+} & = & \frac{M_2}{2\omega_{p2}} \frac{  [ \Delta\Gamma - \frac{i}{16\pi} \Im( h^\dag h_{12})(M_1-M_2)\gamma_5 ](M_2-\omega_{p2}\gamma_0-\vec{p\gamma} )  }{M_1^2-M_2^2-iM_1\Gamma_1+iM_2\Gamma_2} \,, \nonumber \\
  \pN_{2A}^{21+} & = & \frac{M_2}{2\omega_{p2}} \frac{  (M_2-\omega_{p2}\gamma_0-\vec{p\gamma} ) [ \Delta\Gamma - \frac{i}{16\pi} \Im( h^\dag h_{12})(M_1-M_2)\gamma_5 ] }{M_1^2-M_2^2-iM_1\Gamma_1+iM_2\Gamma_2} \,, \nonumber 
\end{eqnarray}
where
\begin{equation}
  \Delta\Gamma \equiv  (M_1+M_2)\frac{\Re( h^\dag h_{12})}{16\pi} \;.
\end{equation}
The \textit{CP} conjugated propagators differ by the sign of $\Im( h^\dag h_{12})$.

Using the vacuum initial condition (\ref{ICLO}) for the statistical propagator
 in leading order in the Yukawa couplings, the coefficients $F_{JI}$ are given by
\begin{eqnarray}
  F_{JI}^\epsilon & = &  f_{FD}(\omega_{p1}) \tr \Big[ \frac{M_1-\vec{p\gamma}}{\omega_{p1}}(\gamma^{11\epsilon}_{JI} - \bar \gamma^{11\epsilon}_{JI} )\Big] \nonumber\\
&& + f_{FD}(\omega_{p2}) \tr \Big[ \frac{M_2-\vec{p\gamma}}{\omega_{p2}}(\gamma^{22\epsilon}_{JI} - \bar \gamma^{22\epsilon}_{JI} )\Big]\;,
\end{eqnarray}
Lets consider for concreteness the first contribution. At leading order in Yukawas,
\begin{eqnarray}
  \gamma^{11\epsilon}_{JI} - \bar \gamma^{11\epsilon}_{JI} & = & -i\gamma_0 \pN^{11\epsilon}_{AJ} P_L \pLH_{\rho\,\vec{p}}^{\epsilon\epsilon} P_R \Delta\pN_{RI}^{\epsilon} \gamma_0 \nonumber\\
 && -i\gamma_0 \Delta\pN_{AJ}^{\epsilon}  P_L \pLH_{\rho\,\vec{p}}^{\epsilon\epsilon} P_R\pN_{RI}^{11\epsilon} \gamma_0 
\end{eqnarray}
where
\begin{eqnarray}
 \Delta\pN_{RI}^{\epsilon} & \equiv & \Re(\yu^\dag\yu)_{12} (\pN^{21\epsilon}_{RI}-\pNcp^{21\epsilon}_{RI} ) +i\Im(\yu^\dag\yu)_{12} (\pN^{21\epsilon}_{RI}+\pNcp^{21\epsilon}_{RI} ) \,, \nonumber\\
 \Delta\pN_{AI}^{\epsilon} & \equiv &  \Re(\yu^\dag\yu)_{12} (\pN^{12\epsilon}_{AI}-\pNcp^{12\epsilon}_{AI} )  -i\Im(\yu^\dag\yu)_{12} (\pN^{12\epsilon}_{AI}+\pNcp^{12\epsilon}_{AI} ) \,.
\end{eqnarray}
From the explicit expressions for $\pN^{ij\epsilon}_{R(A)I}$ given above, one obtains
\begin{eqnarray}
 \Delta\pN_{R1}^{+} & = & i \frac{\Im[(\yu^\dag\yu)_{12}^2]}{16\pi} \frac{M_1}{\omega_{p1}}\frac{\{M_1P_L+M_2P_R\}[M_1-\omega_{p1}\gamma_0-\vec{p\gamma}]}{M_1^2-M_2^2+iM_1\Gamma_1-iM_2\Gamma_2} \,,\nonumber\\
 \Delta\pN_{R1}^{-} & = & i \frac{\Im[(\yu^\dag\yu)_{12}^2]}{16\pi} \frac{M_1}{\omega_{p1}}\frac{\{M_1P_L+M_2P_R\}[M_1+\omega_{p1}\gamma_0-\vec{p\gamma}]}{M_1^2-M_2^2-iM_1\Gamma_1+iM_2\Gamma_2} \,,\nonumber\\
 \Delta\pN_{R2}^{+} & = & -i \frac{\Im[(\yu^\dag\yu)_{12}^2]}{16\pi} \frac{M_2}{\omega_{p2}}\frac{[M_2-\omega_{p2}\gamma_0-\vec{p\gamma}]\{M_1P_L+M_2P_R\}}{M_1^2-M_2^2+iM_1\Gamma_1-iM_2\Gamma_2} \,,\nonumber\\
 \Delta\pN_{R2}^{-} & = & -i \frac{\Im[(\yu^\dag\yu)_{12}^2]}{16\pi} \frac{M_2}{\omega_{p2}}\frac{[M_2+\omega_{p2}\gamma_0-\vec{p\gamma}]\{M_1P_L+M_2P_R\}}{M_1^2-M_2^2-iM_1\Gamma_1+iM_2\Gamma_2} \,,\nonumber
\end{eqnarray}
\begin{eqnarray}
 \Delta\pN_{A1}^{+} & = & i \frac{\Im[(\yu^\dag\yu)_{12}^2]}{16\pi} \frac{M_1}{\omega_{p1}}\frac{[M_1-\omega_{p1}\gamma_0-\vec{p\gamma}]\{M_1P_R+M_2P_L\}}{M_1^2-M_2^2-iM_1\Gamma_1+iM_2\Gamma_2} \,,\nonumber\\
 \Delta\pN_{A1}^{-} & = & i \frac{\Im[(\yu^\dag\yu)_{12}^2]}{16\pi} \frac{M_1}{\omega_{p1}}\frac{[M_1+\omega_{p1}\gamma_0-\vec{p\gamma}]\{M_1P_R+M_2P_L\}}{M_1^2-M_2^2+iM_1\Gamma_1-iM_2\Gamma_2} \,,\nonumber\\
 \Delta\pN_{A2}^{+} & = & -i \frac{\Im[(\yu^\dag\yu)_{12}^2]}{16\pi} \frac{M_2}{\omega_{p2}}\frac{\{M_1P_R+M_2P_L\}[M_2-\omega_{p2}\gamma_0-\vec{p\gamma}]}{M_1^2-M_2^2-iM_1\Gamma_1+iM_2\Gamma_2} \,,\nonumber\\
 \Delta\pN_{A2}^{-} & = & -i \frac{\Im[(\yu^\dag\yu)_{12}^2]}{16\pi} \frac{M_2}{\omega_{p2}}\frac{\{M_1P_R+M_2P_L\}[M_2+\omega_{p2}\gamma_0-\vec{p\gamma}]}{M_1^2-M_2^2+iM_1\Gamma_1-iM_2\Gamma_2} \,,\nonumber
\end{eqnarray}
\begin{eqnarray}
  \pN_{AJ}^{11+} &=& \pN_{RJ}^{11+}=-\frac{i}{2}\delta_{1J}\frac{M_1-\omega_{p1}\gamma_0-\vec{p\gamma}}{\omega_{p1}} \,,\nonumber\\
  \pN_{AJ}^{11-} &=& \pN_{RJ}^{11-}=\frac{i}{2}\delta_{1J}\frac{M_1+\omega_{p1}\gamma_0-\vec{p\gamma}}{\omega_{p1}}\;.
\end{eqnarray}
Finally, we obtain the following result for the coefficients $F_{JI}^\epsilon$:
\begin{eqnarray}
  F_{11}^+ \ =\ F_{11}^- & = & -\frac{ (k\omega_{p1}-\vec{kp}) f_{\lepton\higgs}(k,q)f_{FD}(\omega_{p1}) }{2k\omega_{p1}} \frac{\Im[(\yu^\dag\yu)_{12}^2]}{4\pi}\nonumber\\ 
&& \times\Re\frac{M_1M_2}{M_1^2-M_2^2+ iM_1\Gamma_1- iM_2\Gamma_2} \;.\nonumber
\end{eqnarray}
\begin{eqnarray}
  F_{22}^+ \ =\ F_{22}^- & = & -\frac{ (k\omega_{p2}-\vec{kp}) f_{\lepton\higgs}(k,q)f_{FD}(\omega_{p2}) }{2k\omega_{p2}} \frac{\Im[(\yu^\dag\yu)_{12}^2]}{4\pi}\nonumber\\ 
&& \times\Re\frac{M_1M_2}{M_1^2-M_2^2+ iM_1\Gamma_1- iM_2\Gamma_2} \;, \nonumber
\end{eqnarray}
\begin{eqnarray}
  F_{12}^\pm & = & \frac{(\omega_{p1}+\omega_{p2}) f_{\lepton\higgs}(k,q)}{4k\omega_{p1}\omega_{p2}} \frac{\Im[(\yu^\dag\yu)_{12}^2]}{4\pi}\nonumber\\
&& \times \frac12 \Bigg\{ \frac{M_1M_2(k\omega_{p2}-\vec{kp})f_{FD}(\omega_{p1})}{M_1^2-M_2^2\pm iM_1\Gamma_1\mp iM_2\Gamma_2} + \frac{M_1M_2(k\omega_{p1}-\vec{kp})f_{FD}(\omega_{p2})}{M_1^2-M_2^2\mp iM_1\Gamma_1\pm iM_2\Gamma_2} \Bigg\} \;,\nonumber
\end{eqnarray}
\begin{eqnarray}
  F_{21}^\pm & = & \frac{(\omega_{p1}+\omega_{p2}) f_{\lepton\higgs}(k,q)}{4k\omega_{p1}\omega_{p2}} \frac{\Im[(\yu^\dag\yu)_{12}^2]}{4\pi}\nonumber\\
&& \times \frac12 \Bigg\{ \frac{M_1M_2(k\omega_{p2}-\vec{kp})f_{FD}(\omega_{p1})}{M_1^2-M_2^2\mp iM_1\Gamma_1\pm iM_2\Gamma_2} + \frac{M_1M_2(k\omega_{p1}-\vec{kp})f_{FD}(\omega_{p2})}{M_1^2-M_2^2\pm iM_1\Gamma_1\mp iM_2\Gamma_2} \Bigg\} \;.\nonumber
\end{eqnarray}

\end{appendix}

%\bibliographystyle{mgBibTexStyle}
%\bibliography{references}

\end{document}